  \documentclass[prc,amsmath,amsfonts,showpacs,eqsecnum,floatfix,nofootinbib]{revtex4}

 \usepackage{axodraw}   
 \usepackage{graphicx}  
 \usepackage{pstricks}  
 \usepackage{bm}    

\begin{document}
\hyphenation{Rijken}
\hyphenation{Nijmegen}
 
\title{
    Quark-Quark and Quark-Nucleon Potential Model \\   
    Extended-soft-core Meson-exchange Interactions }
\author{Th.A.\ Rijken}
\affiliation{ Institute of Mathematics, Astrophysics, and Particle Physics \\
 Radboud University, Nijmegen, The Netherlands}               
\author{Y.\ Yamamoto}
\affiliation{Nishina Center for Accelerator-Based Science, Institute for Physical\\
 and Chemical Research (RIKEN), Wako, Saitama, 351-0198, Japan.}

\pacs{13.75.Cs, 12.39.Pn, 21.30.+y}

\date{version of: \today}
 
\begin{abstract}                                       
	The Quark-quark (QQ) and Quark-nucleon (QN) interactions in this paper are derived from 
the Extended-soft-core (ESC) interactions.  The meson-quark-quark (MQQ) vertices 
are determined in the framework of the constituent
quark-model (CQM). These vertices are such that upon folding with the 
ground-state baryonic quark wave functions the one-boson-exchange amplitudes
for baryon-baryon (BB), and in particularly for nucleon-nucleon (NN), 
are reproduced. This opens the attractive 
possibility to define meson-quark interactions at the quark-level which are
directly related to the interactions at the baryon-level. The latter have been
determined by the baryon-baryon data. 
Application of these "realistic" quark-quark interactions in the
quark-matter phase is presumably of relevance for the description of 
highly condensed matter, as e.g. neutron-star matter.

These quark-quark potentials consist of local- and non-local-potentials 
due to (i) One-boson-exchanges 
(OBE), which are the members of nonets of pseudo-scalar-, vector-, scalar-, and
axial-mesons, (ii) Diffractive exchanges, (iii) Two pseudo-scalar exchange (PS-PS),
and (iv) Meson-Pair-exchange (MPE). Both the OBE- and Pair-vertices are regulated 
by gaussian form factors producing potentials with a soft behavior near the origin.
The assignment of the cut-off masses for the BBM-vertices is dependent on the 
SU(3)-classification of the exchanged mesons for OBE, and a similar scheme 
for MPE. The model is presented in the framework of the Kadyshevski formalism,
which has the advantage that in momentum space fully relativistic potentials
can be used in principle.

Like previous ESC models, the recent ESC16 describes nucleon-nucleon (NN), 
hyperon-nucleon (YN), and hyperon-hyperon (YY) 
 in a unified way using broken SU(3)-symmetry. 
Novel ingredients are the inclusion of 
(i) the axial-vector meson potentials, (ii) a zero in the scalar- and axial-vector 
meson form factors. These innovations
made it possible to keep the parameters of the model closely 
to the predictions of the quark-antiquark pair creation (QPC) model,
with a dominance of the $^3P_0$-pair creation.
This is also the case for the flavor SU(3) $F/(F+D)$-ratio's. 
In this QPC-model to the couplings in the framework of the CQM the mesons
are coupled directly to the quarks. Therefore, it is most natural to consider 
meson-exchange on the quark-level as the basis for the meson-exchange
BB-potentials.
In this paper we derive the QQ- and QN-interactions for the two-quark and
	the quark-nucleon channels
of the basic isodoublet  i.e. U,D quarks: (i) UU-, UD-, and DD-channels, and
(ii) UP-, UN- and DN-channels, with P=proton, N=neutron. Applications 
of these potentials can be made for the mixed nuclear and quaurk matter,
which probably occurs for example inside neutron statrs.
 \end{abstract}
 \pacs{13.75.Cs, 12.39.Pn, 21.30.+y}

\maketitle

 \twocolumngrid                                          
\section{Introduction}
\label{sec:1}
In Nucleon-quark mixed matter the nucleons and deconfined quarks interact which each 
other naturally. For a study of such matter, in for example neutron stars, a model
for these interactions is necessary.\\
The Quark-quark and Quark-nucleon/baryon interactions in this paper are derived from 
the Extended-soft-core (ESC) interactions. In \cite{TAR-QQ14,TAR-0014} we have 
determined the meson-quark-quark (QQM) vertices in the framework of the constituent
quark-model (CQM) \cite{Greenberg64,Mic69,LeY73,Man84}. 
These QQM-vertices are such that upon folding with the effective
ground-state baryonic harmonic oscillator quark wave functions, 
the one-boson-exchange amplitudes
for nucleon-nucleon (NN) are reproduced. This opens the attractive 
possibility to define meson-quark interactions at the quark-level which are
directly related to the interactions at the baryon-level. The latter have been
determined by the baryon-baryon data. 
These "unified" quark-quark and quark-nucleon/baryon
interactions can be applied to the 
nucleon-quark mixed-matter phase, which is relevant for the description of 
highly condensed matter, as e.g. neutron-star matter.

In QCD two non-perturbative effects occur: (i) confinement and (ii) chiral
symmetry breaking. The SU(3)$_L$xSU(3)$_R$ chiral symmetry is spontaneously 
broken to an SU(3)$_v$ symmetry at some scale $\Lambda_{\chi SB} \approx 1$ GeV.
Below this scale there is an octet of pseudoscalar Goldstone-bosons: $(\pi, K, \eta)$.
The confinement scale $\Lambda_{QCD} \approx 100-330$ MeV. The complex QCD-vacuum
structure can be described as an BPST instanton/anti-instanton liquid giving the
valence quarks a dynamical or constituent effective mass $\approx M_N/3$ 
\cite{BPST75,DY-PE86}. This corresponds to the
CQM \cite{Man84}, which is the basis for the quark-quark and quark-nucleon interactions 
proposed in this paper.

The QQ-interactions in this paper 
consist of local- and non-local-potentials due to (i) One-boson-exchanges 
(OBE), which are the members of nonets of pseudo-scalar-, vector-, scalar-, and
axial-mesons, (ii) Diffractive exchanges, (iii) Two pseudo-scalar exchange (PS-PS),
and (iv) Meson-Pair-exchange (MPE). Both the OBE- and Pair-vertices are regulated 
by gaussian form factors producing potentials with a soft behavior near the origin.
The assignment of the cut-off masses for the BBM-vertices is dependent on the 
SU(3)-classification of the exchanged mesons for OBE, and a similar scheme 
for MPE.        

The ESC-models in general, and so also the recent version ESC16 \cite{NRY19a,NRY19b,NRY19c},
describe nucleon-nucleon (NN) and hyperon-nucleon (YN) in a unified way using broken 
SU(3)-symmetry. Novel ingredients in ESC16 are the inclusion of (i) the axial-vector
meson potentials, (ii) a zero in the scalar- and axial-vector meson form factors. 
These innovations
made it possible for the first time to keep the parameters of the model closely 
to the predictions of the $^3P_0$ quark-antiquark creation (QPC) model \cite{Mic69,LeY73}.
This is also the case for the $F/(F+D)$-ratio's. 
The application of the QPC model to the couplings was executed in the framework of
the constituent quark-model. Therefore, it is most natural to consider 
meson-exchange on the quark-level. 
In this paper we derive the QQ-interactions for the two-quark channels
of the basic triplet i.e. U,D, and S quarks: (i) UU-, UD-, and DD-, 
(ii) US- and DS-, (iii) SS-channels.

The BBM-vertices are described by coupling constants and
form factors, which correspond to the Regge residues at high
energies \cite{Rij85}. The form factors are taken
to be of the gaussian-type, like the residue functions in many
Regge-pole models for high energy scattering.
Although the gaussian quark wave functions lead to gaussian type of form factors, 
also in (nonrelativistic) quark models (QM's)
a gaussian behavior of the form factors is most natural,
because the mesons are Reggeons. These quark-quark-meson form factors evidently
guarantee a soft behavior of the potentials in configuration
space at small distances.
 
In the ESC models, see {\it e.g.} \cite{RNY10a}, the assignment of the cut-off parameters 
in the form factors is made for the individual baryon-baryon-meson (BBM) vertices, 
constrained by broken $SU(3)$-symmetry. The same scheme we follow here for the
QQM-vertices.

Confinement is related to the infrared behavior of QCD. This plays an important role 
when the quarks are not close together. In quark-matter the quark-density is 
high and therefore the quark-quark interaction is dominantly of short range. 
So, the infrared behavior can be ignored, being the justification for the
use of the same formalism for quarks in (dense) quark matter as for nucleons 
in nuclear matter.

The contents of this paper are as follows. 
In section~\ref{sec:17} we review some facts about the "constituents quarks", 
within the context of spontaneously broken chiral symmetry (Nambu-Goldstone), 
and the complex structure of the QCD vacuum. 
In section~\ref{sec:match} the relation between the ESC BBM-couplings and the 
QQM-couplings is argued for the CQM.
In section~\ref{sec:2} we present the Kadyshevsky formalism \cite{Kad64,Kad67,Kad68,Kad70}
for baryon-baryon channels and the three-dimensional "quasi-potential equation.
From this a in section~\ref{sec:4} a 
Thompson-type relativistic Lippmann-Schwinger is derived using a standard \cite{Kad64}
on-energy-shell version of the Kadyshevsky equation.
In section~\ref{sec:15} the Lippmann-Schwinger and the Bethe-Goldstone equations are given.
In section~\ref{app:22} the ESC meson-quark-quark and meson-nucleon-nucleon
interaction Hamiltonians are displayed, both for the
QQM/NNM-vertices as well as for the pair-vertices QQMM/NNMM.
Here also the meson-pair interaction Hamiltonians are given in
the context of $SU(3)$. Expressions for the meson-pair-exchange (MPE)
graphs are given, again in an immediately programmable form.
In section~\ref{sec:12} we describe 
the $S=0,-1,-2$ $QQ$-channels on the isospin (i) and hypercharge (y) and particle 
basis. Here also the SU(3)-structure of the QQM/NNM-couplings are given both in the
$3\times 3$-matrix and cartesian-nonet description. 
For the calculation of the SU(3) factors of the OBE, TME and MPE diagrams we refer to
\cite{Rij04a,Rij04b}.
Here also the gaussian form factors are mentioned.
In section~\ref{sec:5} the connection between the QQM- and BBM-couplings are listed.
In section~\ref{sec:conf} the one-gluon-exchange (OGE) and confining potentials are described.
In section~\ref{sec:njl3} the SU(3) Nambu-Jona-Lasinio (NJL) form of the instanton potential is
worked out. 
with the ESC-couplings is discussed.
In section~\ref{sec:15a} the simultaneous $NN \oplus YN$ fitting procedure of the
meson-exchange parameters is briefly reviewed, and 
the results for the coupling constants and $F/(F+D)$-ratios
for OBE and MPE are given. 
Furthermore, results for the QQ-, QN-, and NN-phase-shifts are given. 
{\it This is only meant for a comparison of the strength of the QQ- and 
QN-interactions and the NN-interactions.
}
In section~\ref{sec:discussion} a summary and an outlook is given.\\
 In Appendix~\ref{app:kadgmat} the the Bethe-Goldstone-Kadyshevsky equation and the
 correspondent G-matrix are described.
 In Appendix~\ref{app:qpc} a simple model for the relation between the 
 meson-couplings using the Fierz-transformation is described.
In Appendix~\ref{app:MQV} the complete meson-quark vertices in Pauli-spinor
space are given.
In Appendix~\ref{app:OBE} the one-boson-exchange quark-quark potentials in 
momentum- and configuration-space are given for the vertices which
also occur at the baryon-level.
In Appendix~\ref{app:OBE2} the additional quark-quark potentials are given, which are
due to the extra meson-vertices at the quark-level.
 Next we included some miscellaneous topics:
 In Appendix~\ref{app:MMQQ} discusses the inclusion of the Z-graphs in the
  MPE-interaction is implicit.

\section{Constituent Quarks and Instantons}
\label{sec:17}
The spectra of the nucleons, $\Delta$ resonances and the hyperons $\Lambda, \Sigma, \Xi$ are
descibed in detail by the Glozman-Riska model \cite{Gloz96a}.
This is a modern version of the CQM \cite{Greenberg64} 
based on the Nambu-Goldstone spontaneous chiral-symmetry breaking (SCSB) 
with quarks interacting by the exchange of the SU(3)$_F$ octet of pseudoscalar
mesons \cite{Gloz96a}. The pseudoscalar octet are the Nambu-Goldstone bosons (NGB's) 
associated with the hidden (approximate) chiral symmetry of QCD.
The confining potential is chosen to be harmonic, as is rather common in 
constituent quark models. 
In line with this, we used harmonic wave functions in the derivation of
the connection between the meson-baryon and meson-quark couplings \cite{TAR-QQ14}.
The $\eta'$, which is dominantly an SU(3) singlet, decouples from the original nonet 
because of the U(1) anomaly \cite{Weinberg75,GtH76}. 
According to the two-scale picture of Manohar and Georgi \cite{Man84} the effective
degrees for the 3-flavor QCD at distances beyond that of SCSB 
($\Lambda_{\chi SB}^{-1} \approx 0.2-0.3$ fm),
but within that of the confinement scale $\Lambda_{QCD}^{-1} \approx 1$ fm,
should be the constituent quarks and chiral meson fields.
The two non-perturbative effects in QCD are confinement and chiral symmetry
breaking. The SU(3)$_L\otimes$SU(3)$_R$ chiral symmetry is sponteneously broken to 
an SU(3)$_v$ symmetry at a scale $\Lambda_{\chi SB} \approx 1$ GeV. 
The confinement scale is $\Lambda_{QCD} \approx 100-300$ MeV, which roughly
corresponds to the baryon radius $\approx$ 1 fm.
Due to the complex structure of the QCD vacuum, which can be understood as a
liquid of BPST instantons and anti-instantons \cite{BPST75,DY-PE84a,DY-PE84b,DY-PE86}, 
the valence quarks acquire a dynamical or constituent mass \cite{Weinberg75,Man84,
SHUR84,DY-PE84b,DY-PE86}. 
The interaction between the instanton and the anti-instanton is a dipole-interaction
\cite{SHUR82a}, similar to ordinary molecules: weak attraction at large distances
and strong repulsion at small ones. With the empirical value of the gluon
condensate \cite{SVZ79} as input the instanton density and radius become 
\cite{SHUR82a} $n_c= 8\cdot 10^{-4}\ {\rm GeV}^{-4}$, and 
$\rho_c= (600\ {\rm MeV})^{-1} \approx 0.3\ {\rm fm}$ respectively.
Also, with these parameters the non-perturbative vacuum expectation value for 
the quark fields is
$\langle vac|\bar{\psi}\psi|vac\rangle \approx -10^{-2}\ {\rm GeV}^3$ and 
the quark effective (u,d) masses $\approx 200$ MeV, i.e. 
much larger than the almost massless "current masses".
In the calculation of light quarks in the 
instanton vacuum \cite{DY-PE86}  the effective quark mass $m_Q(p=0)= 345$ MeV was 
calculated, which is remarkably close to the constituent mass $M_N/3$.

Very notable is the role of the instantons for the light meson spectrum. They give a
non-perturbative gluonic interaction between quarks in QCD. For example the instanton-induced 
interaction, as proposed by 't Hooft \cite{GtH76}, generates at low momenta 
the constituent quark mass \cite{DY-PE86}, i.e. breaks chiral symmetry. 
This interaction supplies 
a strong attractive attraction in the pseudoscalar-isovector quark-antiquark
system - pions -, which makes them anomalously light, with zero mass in the 
chiral limit. This is the mechanism by which the pions, being  quark-antiquark 
bound states, appear as Nambu-Goldstone bosons of the SCSB symmetry. 
This strongly attractive interaction is absent 
in vector mesons \cite{Gloz00a,Glozman99}, making the masses of the vector mesons 
$\approx 2m_Q$ in
accordance with $m_\rho \approx m_\omega \approx 2 m_Q$. Since $\alpha_s \approx 0.3$
the one-gluon-exchage (OGE) is weak, and therefore the $\pi-\rho$ mass splitting 
is not due to the perturbative color-magnetic spin-spin interaction between 
the quark and antiquark \cite{Glozman99}. 
Besides explaining the $\pi-\rho$ mass difference, the 
't Hooft interaction also in a natural way solves the $U_A(1)$ problem, and 
gives the reason why the $\eta'$ is heavy, as distinct to the NGB pseudoscalar octet.

The 't Hooft four-fermion instanton mediated interaction  
for the light flavor doublet $\psi=(u,d)$, in the form of a generalized 
Nambu-Jona-Lasinio Lagrangian \cite{NJL61}, is 
\begin{equation}
 {\cal L}_I = G_I\left[(\bar{\psi}\psi)^2-(\bar{\psi} \gamma_5\bm{\tau}\psi)^2
 -(\bar{\psi}\bm{\tau}\psi)^2+(\bar{\psi}\gamma_5\psi)^2\right].
\label{eq:inst1}\end{equation}
Here, the strength of the interaction $G_I$ and the ultraviolet cut-off
scale $1/r_0$ are related in the instanton liquid model \cite{Diakonov96}.
In this model $G_I= \lambda_{ud}/4 = 2n_+/\langle\bar{\psi}\psi\rangle)^2$.
In \cite{Glozman00b} Glozman and Varga show that the t-channel iteration 
of the instanton interaction (\ref{eq:inst1}) leads to isoscalar and isovector
pseudoscalar and scalar exchange quark-quark potentials. Since these potentials 
are already included in our model, the four-fermion instanton interaction 
does not lead to extra potentials. 

In this paper we extend the meson-exchange between quarks by proposing to include,
besides the pseudoscalar, all meson nonets: vector, axial-vector, scalar etc.
{\it Since all these meson nonets can be considered as quark-antiquark bound states,
there is no reason to exclude any of these mesons from the quark-quark interactions.}
{\it Furthermore, our preferred value for the constituent quark mass has 
a solid basis in the instanton-liquid model of the QCD vacuum.}

\onecolumngrid
\begin{flushleft}
\rule{16cm}{0.5mm}
\end{flushleft}

\section{ESC-potentials and the Constituent Quark-model}              
\label{sec:match} 
The fitted ESC16-couplings and the QPC-couplings agree very well as shown in \cite{NRY19a}.
In particular, the SU(6)-breaking improves the agreement significantly. 
The calculation of Table II in Ref.~\cite{NRY19a} uses the constituent quark model (CQM) in the 
SU(6)-version of \cite{LeY73}. In Appendix~\ref{app:qpc} a simple model for the 
quark-antiquark creation process exhibits the main features of the meson-coupling pattern 
in the ESC models. 
Since such calculations implicity uses the direct coupling of
the mesons to the quarks, it defines the QQM-vertex. Then, OBE-potentials can be 
derived by folding meson-exchange with the quark wave functions of the baryons.
Prescribed by the Dirac-structure, at the baryon level the vertices 
have in Pauli-spinor space the 1/M$_B$-expansion
\begin{eqnarray}
 \bar{u}(p',s') \Gamma u(p,s) &=& \chi_{s'}^{\prime \dagger}\left\{
 \Gamma_{bb}+\Gamma_{bs}\frac{\bm{\sigma}\cdot{\bf p}}{E+M} 
 -\frac{\bm{\sigma}\cdot{\bf p}'}{E'+M'} \Gamma_{sb} 
 -\frac{\bm{\sigma}\cdot{\bf p}}{E+M} \Gamma_{ss} 
 \frac{\bm{\sigma}\cdot{\bf p}'}{E'+M'} \Gamma_{sb}\right\}\ \chi_s \nonumber\\
 &\equiv& \sum_l c_{BB}^{(l)}\ \left[\chi_{s'}^{\prime \dagger}O_l({\bf p}',{\bf p})\
 \chi_s\right]\ (\sqrt{M'M})^{\alpha_l}\ \ (l=bb,bs,sb,ss).
\label{eq:CQM.1}\end{eqnarray}
This expansion is general and does not depend on the internal structure of the
baryon. A similar expansion can be made on the quark-level, but now with quark masses $m_Q$ and
coefficients $c_{QQ}^{(l)}$.
It appears that in the CQM, i.e. $m_Q=M_B/3$, the QQM-vertices can be 
chosen such that the ratio's $c_{QQ}^{(l)}/c_{BB}^{(l)}$ are constant for each type of
meson \cite{TAR-QQ14}  Then, these coefficients can be made equal by (i) scaling the
couplings, (ii) introducing some extra couplings at the quark level, and (iii) introducing
a QQM gaussian form factor.
Ipso facto this defines a meson-exchange quark-quark interaction.                     

\begin{flushleft}
\rule{16cm}{0.5mm}
\end{flushleft}
\section{Kadyshevsky Equations in Momentum Space}
\label{sec:2}
We envisage the interaction between two (constituent) quarks in a dense
medium of baryons and/or quarks. In such a condition it is appropriate
to consider the QQ-correlations in the G-matrix formalism in the setting of
the Bethe-Goldstone equations \cite{BG57,BG71}. To make contact with the
3-dimensional potental formalism we employ the Kadyshevsky formalism \cite{Kad64}.

\subsection{Relativistic Two-Body Equation}
\label{sec:3}
We consider the quark-quark and quark-nucleon baryon-baryon reactions
\begin{subequations}\label{eq:3.1}
\begin{eqnarray}
    Q_{a}(p_{a},s_{a})+Q_{b}(p_{b},s_{b}) \rightarrow
    Q_{a}'(p_{a}',s_{a}')+Q_{b}'(p_{b}',s_{b}'), \\
    Q_{a}(p_{a},s_{a})+N_{b}(p_{b},s_{b}) \rightarrow
    Q_{a}'(p_{a}',s_{a}')+N_{b}'(p_{b}',s_{b}').
\end{eqnarray}\end{subequations}
The set-up of the scattering formalism for baryon-baryon interactios is
given in \cite{NRY19a,NRY19b,NRY19c} and is given here for completeness.

\noindent Introducing, as usual, the total and relative four-momentum for the initial
and final state
\begin{equation}\begin{array}{lcl}
 P = p_{a} + p_{b} &,& P' = p_{a}' + p_{b}'\ , \\
 p = \frac{1}{2}(p_{a}-p_{b}) &,& p' = \frac{1}{2}(p_{a}'-p_{b}')\ ,
\end{array} \label{eq:3.2}\end{equation}
We use in the following the notation $P_0 \equiv W$ and $P_0' \equiv W'$.
In the Kadyshevsky formulation one introduces four-momenta spurions,
making formally four-momentum conservation at the vertices. These are 
described by quasi-particle states $|\kappa\rangle$, normalized by
$\langle \kappa' | \kappa\rangle = \delta(\kappa'-\kappa)$. Then the 
four-momentum of such a state is $\kappa n^\mu$, where $n^\mu$ is 
time-like with $n^0 > 0$ and $ n^2=1$. 
So, we consider the process in (\ref{eq:3.1})
with non-conservation of the four-momentum, i.e. off-momentum-shell.
This off-shellness is given by 
\begin{equation}
   p_a+p_b + \kappa n = p_a' + p_b' + \kappa' n
\label{eq:3.3}\end{equation}
In the following, the on-mass-shell momenta for the initial
and final states are denoted respectively by $p_{i}$ and $p_{f}$.
So, $p_{i 0}=E({\bf p}_i)=\sqrt{{\bf p}_i^{2}+M^{2}}$ and
$p_{f 0}=E({\bf p}_f)=\sqrt{{\bf p}_f^{2}+M^{2}}$.

\begin{figure}   
 \begin{center}
 \resizebox{7.25cm}{!}        
 {\includegraphics[225,650][425,900]{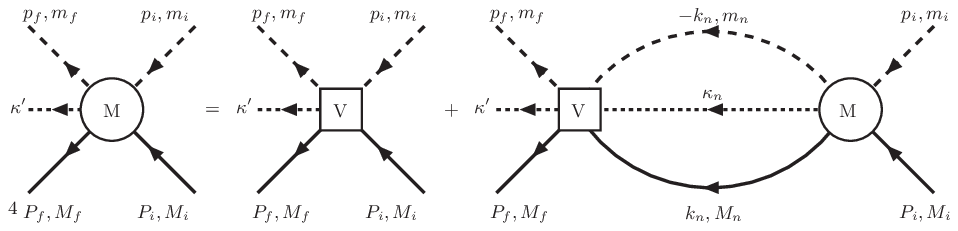}}
  \caption{\sl M-matrix: Kadyshevsky-Integral Equation}                         
  \label{fig:inteq5} 
 \end{center}
 \end{figure}

In the Kadyshevsky-formulation the particles are on-mass-shell in the 
Green-functions. The on-mass-shell propagator $S^{(\pm)}(p)$ of a 
spin-0 particle can be written as 
\begin{equation}
   S^{(\pm)}(p)= 
  \delta_{\pm}(p^2-M^2) = \frac{1}{2E({\bf p})}  
   \delta\left(p_{0} \mp E({\bf p})\right)\ ,
\label{eq:3.5}\end{equation}
with $\delta_{\pm}(p^2-M^2) \equiv \theta(\pm p^0)\delta(p^2-M^2)$.  
The propagator $G_0(\kappa)$ for the quasi-particles is given by 
\cite{Kad67}
\begin{equation}
 G_0(\kappa) = (1/2\pi) \left[1/(\kappa-i\delta)\right]\ .
\label{eq:3.6}\end{equation}
In the Kadyshevsky-formalism the rules for the computation of the 
off-shell S-matrix, denoted by $R$, corresponding to the analogs of
the Feynman graphs are given \cite{Kad67,Kad68,Kad70}. 
\footnote{
For a general field-theoretical treatment of the Kadyshevsky approach to relativistic
two-body scattering, see Refs. \cite{RW09,RW10}.}
We introduce the usual $M$-matrix by
\begin{eqnarray}
 R_{\kappa',\kappa}(p_a',p_b'; p_a, p_b) &=&
 \delta(\kappa'-\kappa)\delta(p_a'-p_a)\delta(p_b'-p_b) - (2\pi)^4 i
 \delta^4(\kappa' n + p_a' + p_b'- p_a - p_b - \kappa n)\cdot
 \nonumber\\ && \times 
 M_{\kappa',\kappa}(p_a',p_b'; p_a, p_b)\ .
\label{eq:3.7}\end{eqnarray}
Notice that the $S$-matrix is given by $R_{0,0}$ \cite{Kad67}. 
We also observe that
\begin{equation}
 \delta(\kappa'-\kappa)\delta(p_a'-p_a)\delta(p_b'-p_b) =            
 \delta(P'+\kappa' n - P - \kappa n)     
 \delta(p_a'-p_a)\delta(p_b'-p_b)\ ,           
\label{eq:3.8}\end{equation}
showing the overall 4-momentum conservation for the $R$-matrix, including
the momentum spurions.\\
The M-amplitudes satisfy the Kadyshevsky equation              
\begin{eqnarray}
M_{\kappa',\kappa}(p_a',p_b'; p_a, p_b) &=& 
I_{\kappa',\kappa}(p_a',p_b'; p_a, p_b) + 
\int d^4p_a^{\prime\prime}\int d^4p_b^{\prime\prime}\int d\kappa^{\prime\prime}
I_{\kappa'\kappa^{\prime\prime}}(p_a',p_b'; p_a^{\prime\prime}, 
p_b^{\prime\prime})\cdot
 \nonumber\\ && \times
G_{\kappa^{\prime\prime}}(p_a^{\prime\prime}, 
p_b^{\prime\prime}) M_{\kappa^{\prime\prime},\kappa}
(p_a^{\prime\prime},p_b^{\prime\prime}; p_a, p_b)\cdot \delta(
p_a^{\prime\prime}+p_b^{\prime\prime}+
\kappa^{\prime\prime}n-p_a-p_b-\kappa n),  
\label{eq:3.9}\end{eqnarray}
which is displayed in Fig.~\ref{fig:inteq5}.
Here the propagation of the two nucleons and of the quasi-particle is
described by
\begin{eqnarray}
G_\kappa(p_a,p_b)_{\alpha',\beta';\alpha,\beta} &=& \frac{-1}{(2\pi)^2}     
 \delta(p_a^2-M_a^2) \delta(p_b^2-M_b^2)\cdot G_0(\kappa)\ .
\label{eq:3.10}\end{eqnarray}

\section{Three-dimensional Two-Body Equations}   
\label{sec:4}
The Kadyshevsky analog (\ref{eq:3.9}) 
of the Bethe-Salpeter equation we write in the form
\begin{eqnarray}
 && M_{\kappa',\kappa}(p_a',p_b'; p_a, p_b) = 
 I_{\kappa',\kappa}(p_a',p_b'; p_a, p_b) +   
\int d^4p_a^{\prime\prime}\int d^4p_b^{\prime\prime}\int d\kappa^{\prime\prime}
 \cdot \nonumber\\ && \times
I_{\kappa',\kappa^{\prime\prime}}(p_a',p_b'; p_a^{\prime\prime}, 
p_b^{\prime\prime})
G_{\kappa^{\prime\prime}}(p_a^{\prime\prime},p_b^{\prime\prime}) 
M_{\kappa^{\prime\prime},\kappa}(p_a^{\prime\prime},p_b^{\prime\prime};        
p_a, p_b)\cdot \nonumber\\ && \times
\delta(p_a^{\prime\prime}+p_b^{\prime\prime}+
\kappa^{\prime\prime}n-p_a-p_b-\kappa n)\ .
\label{eq:4.1}\end{eqnarray}
In the CM-frame we have
 \begin{equation}
  P = ( W , {\bf 0}) \hspace{0.2cm} , \hspace{0.2cm} p = ( 0 , {\bf p})
  \hspace{0.2cm} ; P' = ( W' , {\bf 0}) \hspace{0.2cm} , 
  \hspace{0.2cm} p' = ( 0 , {\bf p}')\ .
\label{eq:4.2}\end{equation}
Following \cite{Kad67,Kad70} we assume that the unit vector $n^\mu$, 
which defines the time axis, is collinear to $P=p_a+p_b$ and hence also
to $P'=p_a'+p_b'$. 
Then
\footnote{Notice that with this choice for $n^\mu$, the four-velocity of
the system is conserved even off the energy-shell.} 
\begin{equation}
 n^\mu = \frac{p_a^\mu+p_b^\mu}{\sqrt{(p_a+p_b)^2}} =
 \frac{p_a^{\prime\mu}+p_b^{\prime\mu}}{\sqrt{(p'_a+p'_b)^2}}
 \stackrel{CM}{\longrightarrow} (1,{\bf 0})\ .
\label{eq:4.3}\end{equation}
In the CM-variables, equation (\ref{eq:4.1}), for the $(+,+)$-components only,
reads
\begin{eqnarray}
 && M_{\kappa',\kappa}(p',W';p,W) = I_{\kappa',\kappa}(p',W';p,W) +
\int dW^{\prime\prime}\int d^4p^{\prime\prime}\int d\kappa^{\prime\prime}
 \cdot \nonumber\\ && \times
I_{\kappa',\kappa^{\prime\prime}}(p',W'; p^{\prime\prime}, 
W^{\prime\prime})
G_{\kappa^{\prime\prime}}(p^{\prime\prime},W^{\prime\prime}) 
M_{\kappa^{\prime\prime}}(p^{\prime\prime},W^{\prime\prime}; p, W)
\cdot \nonumber\\ && \times
\delta\left[W^{\prime\prime}-W+(\kappa^{\prime\prime}-\kappa)n_0\right]\ .
\label{eq:4.4}\end{eqnarray}
In the CM-frame, the two-nucleon propagator (\ref{eq:3.10}) becomes
\begin{eqnarray}
G_\kappa(W^{\prime\prime},p^{\prime\prime}) &=&
 \frac{-1}{(2\pi)^2}     
 \delta\left(\frac{1}{2}W^{\prime\prime}+p_0^{\prime\prime}-
 E_a^{\prime\prime}\right)
 \delta\left(\frac{1}{2}W^{\prime\prime}-p_0^{\prime\prime}-
 E_b^{\prime\prime}\right) G_0(\kappa^{\prime\prime})\ .
\label{eq:4.5}\end{eqnarray}
Now, the integrations over $W^{\prime\prime}, p_0^{\prime\prime}$, and 
$\kappa^{\prime\prime}$ can be carried through in (\ref{eq:4.4}) giving
\begin{eqnarray}
 && M_{\kappa',\kappa}({\bf p'},W';{\bf p},W) = 
I_{\kappa',\kappa}({\bf p'},W';{\bf p},W) +
\int \frac{d^3p^{\prime\prime}}{(2\pi)^3}
 \cdot \nonumber\\ && \times
I_{\kappa',\kappa^{\prime\prime}}({\bf p'},W'; {\bf p}^{\prime\prime}, 
W^{\prime\prime})
\left(\frac{M_aM_b}{E_a^{\prime\prime}E_b^{\prime\prime}}\right)
\frac{1}{\sqrt{s^{\prime\prime}}-\left(\sqrt{s}+\kappa\right)- i\epsilon}
M_{\kappa^{\prime\prime},\kappa}
({\bf p}^{\prime\prime},W^{\prime\prime};{\bf p}, W)\ ,
\label{eq:4.6}\end{eqnarray}
with the constraints 
\begin{eqnarray}
 W &=& \sqrt{s}\ \ , \ \ 
 W^{\prime} = \sqrt{s'} =\sqrt{s}+\kappa-\kappa'\ ,\ 
 W^{\prime\prime} = \sqrt{s^{\prime\prime}}= 
E_a^{\prime\prime}+E_b^{\prime\prime}\ .             
\label{eq:4.7}\end{eqnarray}
We notice that the left-half-off-shell $M$-matrix satisfies an integral 
equation of the type
\[
 M_{\kappa',0} = I_{\kappa',0} + \int\ I_{\kappa',\kappa^{\prime\prime}}\
 G_{\kappa^{\prime\prime}}\ M_{\kappa^{\prime\prime},0}
\]
where the $\kappa$'s are all fixed in terms of the momenta of the 
particles, since
\[
 \kappa^{\prime} = \sqrt{s}-\sqrt{s'}\ \ ,\ \
 \kappa^{\prime\prime} = \sqrt{s}-\sqrt{s^{\prime\prime}}\ .
\]


Defining the $T$-matrix etc. in terms of the left-half-off-shell $M$-matrix 
, and the quasi-potential $K$ in terms of the both left and right off-shell 
interaction kernel $I$, by
\begin{eqnarray}
 T({\bf p'},{\bf p};W) &=& M_{\kappa',\kappa=0}({\bf p'},W';{\bf p},W)\ ,\ 
 K({\bf p'},{\bf p};W) = I_{\kappa',\kappa=0}({\bf p'},W';{\bf p},W)\ , 
\label{eq:4.8}\end{eqnarray}
we will have, instead of (\ref{eq:4.6}),
\begin{eqnarray}
 T({\bf p'},{\bf p};W) &=& K({\bf p'},{\bf p};W) +         
 \int \frac{d^3p^{\prime\prime}}{(2\pi)^3}\
 K({\bf p'},{\bf p}^{\prime\prime};W)\        
\left(\frac{M_aM_b}{E_a^{\prime\prime}E_b^{\prime\prime}}\right)
\frac{1}{\sqrt{s^{\prime\prime}}-\sqrt{s}}\
T({\bf p}^{\prime\prime},{\bf p};W),
\label{eq:4.9}\end{eqnarray}
which is the so-called 'quasi-potential' equation. The quantity K playing the role of a potential
is in general a complicated function of the energy W and is called a 'quasi-potential'.
Notice, that for $\kappa=0$, one has $\kappa'= \sqrt{s}-\sqrt{s'}$, 
and so $\kappa'$ is fixed by $p=|{\bf p}|$ and $p'=|{\bf p}'|$.\\

For equal masses, i.e. $M_a=M_b=M$, we have
\begin{eqnarray}
 E_a^{\prime\prime} &=& E_b^{\prime\prime} = E({\bf p}^{\prime\prime})\ ,
 s = 4 E^2({\bf p)} = 4(p^2+M^2)\ , 
 s^{\prime\prime} = 4 E^2({\bf p}^{\prime\prime}) = 
 4(p^{\prime\prime 2}+M^2).
\label{eq:4.10}\end{eqnarray}
Then, (\ref{eq:4.9}) goes over into the equation
\begin{eqnarray}
 T({\bf p'},{\bf p};W) &=& K({\bf p'},{\bf p};W) +         
 \frac{1}{(2\pi)^3}\int \frac{d^3p^{\prime\prime}}
 {2E({\bf p}^{\prime\prime})}\
 K({\bf p'},{\bf p}^{\prime\prime};W)\        
\frac{M^2}{E({\bf p}^{\prime\prime})\left]E({\bf p}^{\prime\prime})-
 E({\bf p})-i\epsilon\right]}\
T({\bf p}^{\prime\prime},{\bf p};W),
\label{eq:4.11}\end{eqnarray}
which is the quasi-potential equation of Kadyshevsky, see \cite{Kad68}
equation (3.33).

In Appendix~\ref{app:kadgmat} the Bethe-Goldstone-Kadyshevsky equation 
and the corresponding "relativistic" G-matrix are given.
\section{Lippmann-Schwinger and Bethe-Goldstone Equation}
\label{sec:15}
The Lippmann-Schwinger amplitude is obtained from (\ref{eq:4.11}) by the
transformation
\begin{eqnarray}
 {\cal T}({\bf p}',{\bf p}) = N({\bf p}')\ T({\bf p}',{\bf p})\ N({\bf p})\ ,\
 {\cal V}({\bf p}',{\bf p}) = N({\bf p}')\ K({\bf p}',{\bf p})\ N({\bf p}),    
\label{eq:15.1}\end{eqnarray}
 with $N({\bf p}) = M/(\sqrt{2}E({\bf p}))$.
Then, the non-relativistic Lippmann-Schwinger equation is obtained by using
in the Green-function and the potential the non-relativistic approximation
$E({\bf p}) \approx M+{\bf p}^2/2M$ giving
\begin{eqnarray}
 {\cal T}({\bf p'},{\bf p}) &=& {\cal V}({\bf p'},{\bf p}) +         
 \frac{1}{(2\pi)^3}\int \frac{d^3p^{\prime\prime}}
 {2E({\bf p}^{\prime\prime})}\
 {\cal V}({\bf p'},{\bf p}^{\prime\prime})\        
\frac{M}{\left({\bf p}^{\prime\prime 2}-
 {\bf p}^2-i\epsilon\right)}\
 {\cal T}({\bf p}^{\prime\prime},{\bf p}).
\label{eq:15.2}\end{eqnarray}
For the details of the formalism of spin 1/2-1/2 scattering, using the expansion 
in Pauli-invariants, we refer to the papers of the ESC-model e.g. \cite{Rij04a,Rij04b}.         

\twocolumngrid

 \twocolumngrid

The corresponding Bethe-Goldstone equation reads 
\begin{eqnarray}
    G({\bf p}',{\bf p}) &=& V({\bf p}',{\bf p}) +
    \int \frac{d^{3}p''}{(2\pi)^3}\ V({\bf p}',{\bf p}'')\cdot
  \nonumber\\ && \times 
  Q_P({\bf p}^{\prime\prime};p_F)\ 
 g({\bf p}''; W)\; G({\bf p}'',{\bf p})
\label{eq:30.26} \end{eqnarray}
with the standard Green function and Pauli projection operator
\begin{equation}
    g({\bf p};W) = 
    \frac{M_n}{{\bf p}_i^{2}-{\bf p}^{2}+i\delta},\ 
  Q_P({\bf p}^{\prime\prime};p_F)= 1-n_F({\bf p}^{\prime\prime}).
\label{eq:30.27} \end{equation}
The corrections to the approximation $E_{2}^{(+)} \approx g({\bf p}; W)$ 
are of order $1/M^{2}$, which we neglect henceforth.
 
The transition from Dirac-spinors to
Pauli-spinors, is given in Appendix C of Ref.~\cite{Rij91}, where we write for the 
the Bethe-Goldstone equation in the 4-dimensional Pauli-spinor space
\begin{eqnarray}
 {\cal G}({\bf p}',{\bf p})&=&{\cal V}({\bf p}',{\bf p}) + 
 \int \frac{d^{3} p''}{(2\pi)^3}\ {\cal V}({\bf p}',{\bf p}'')\cdot
  \nonumber\\ && \times 
  Q_P({\bf p}^{\prime\prime};p_F)\ 
 g({\bf p}''; W)\ {\cal G}({\bf p}'',{\bf p})\ .
 \label{eq:30.28} \end{eqnarray}

The ${\cal G}$-operator in Pauli spinor-space is defined by
\begin{eqnarray}
 && \chi^{(a)\dagger}_{\sigma'_{a}}\chi^{(b)\dagger}_{\sigma'_{b}}\; 
 {\cal G}({\bf p}',{\bf p})\;
 \chi^{(a)}_{\sigma_{a}}\chi^{(b)}_{\sigma_{b}}  =              
\nonumber\\ && 
 \bar{u}_{a}({\bf p}',\sigma'_{a})\bar{u}_{b}(-{\bf p}',\sigma'_{b})\
 \tilde{G}({\bf p}',{\bf p})\; u_{a}({\bf p},\sigma_{a}) u_{b}(-{\bf p},\sigma_{b}).
\nonumber\\
 \label{eq:30.29} \end{eqnarray}
and similarly for the ${\cal V}$-operator.
Like in the derivation of the OBE-potentials \cite{NRS77,NRS78,MRS89}
we make the off-shell and on-shell the approximation,
  $ E({\bf p})= M + {\bf p}^{2}/2M $
 and $ W = 2\sqrt{{\bf p}_i^{2}+M^{2}} = 2M + {\bf p}_i^{2}/M$ ,     
everywhere in the interaction kernels, which, of course,
is fully justified for low energies only. 
In contrast to these kinds of approximations, of course the full
${\bf k}^{2}$-dependence of the form factors is kept
throughout the derivation of the TME. 
Notice that the gaussian form factors suppress the high momentum
transfers strongly. This means that the contribution to the potentials
from intermediate states which are far off-energy-shell can not
be very large. 

Because of rotational invariance and parity conservation, the ${\cal G}$-matrix, which is
a $4\times 4$-matrix in Pauli-spinor space, can be expanded 
into the following set of in general 8 spinor invariants, see for example 
Ref.~\cite{SNRV71}. Introducing \cite{notation1}
\begin{equation}
  {\bf q}=\frac{1}{2}({\bf p}'+{\bf p})\ , \
  {\bf k}={\bf p}'-{\bf p}\ , \           
  {\bf n}={\bf p}\times {\bf p}',
\label{eq:30.30} \end{equation}
with, of course, ${\bf n}={\bf q}\times {\bf k}$,
we choose for the operators $P_{j}$ in spin-space
\begin{eqnarray}
&&  P_{1}=1,  \hspace{3mm} P_{2}= 
 \mbox{\boldmath $\sigma$}_1\cdot\mbox{\boldmath $\sigma$}_2,
 \nonumber\\[0.0cm]
&& P_{3}=(\mbox{\boldmath $\sigma$}_1\cdot{\bf k})(\mbox{\boldmath $\sigma$}_2\cdot{\bf k})
 -\frac{1}{3}(\mbox{\boldmath $\sigma$}_1\cdot\mbox{\boldmath $\sigma$}_2)
  {\bf k}^2,
 \nonumber\\[0.0cm]
&& P_{4}=\frac{i}{2}(\mbox{\boldmath $\sigma$}_1+
 \mbox{\boldmath $\sigma$}_2)\cdot{\bf n}, \hspace{3mm} 
 P_{5}=(\mbox{\boldmath $\sigma$}_1\cdot{\bf n})(\mbox{\boldmath $\sigma$}_2\cdot{\bf n}),
 \nonumber\\[0.0cm]
 && P_{6}=\frac{i}{2}(\mbox{\boldmath $\sigma$}_1-\mbox{\boldmath $\sigma$}_2)\cdot{\bf n}, 
 \nonumber\\[0.0cm]
 && P_{7}=(\mbox{\boldmath $\sigma$}_1\cdot{\bf q})(\mbox{\boldmath $\sigma$}_2\cdot{\bf k})
 +(\mbox{\boldmath $\sigma$}_1\cdot{\bf k})(\mbox{\boldmath $\sigma$}_2\cdot{\bf q}),
 \nonumber\\[0.0cm]
&& P_{8}=(\mbox{\boldmath $\sigma$}_1\cdot{\bf q})(\mbox{\boldmath $\sigma$}_2\cdot{\bf k})
 -(\mbox{\boldmath $\sigma$}_1\cdot{\bf k})(\mbox{\boldmath $\sigma$}_2\cdot{\bf q}).
\label{eq:30.31} \end{eqnarray}
Here we follow Ref.~\cite{MRS89}, where in contrast to Ref.~\cite{NRS78},
we have chosen $P_{3}$ to be a purely `tensor-force' operator.
The expansion in spinor-invariants reads
\begin{equation}
 {\cal G}({\bf p}',{\bf p}) = \sum_{j=1}^8\ \widetilde{G}_j({\bf p}^{\prime 2},{\bf p}^2,
 {\bf p}'\cdot{\bf p})\ P_j({\bf p}',{\bf p})\ .
\label{eq:30.32} \end{equation}
Similarly to (\ref{eq:30.32}) we expand the potentials $V$. 
In the case of the axial-vector meson exchange there will occur terms
proportional to
\begin{equation}
 P_5'=(\mbox{\boldmath $\sigma$}_1\cdot{\bf q})(\mbox{\boldmath $\sigma$}_2\cdot{\bf q})
 -\frac{1}{3}(\mbox{\boldmath $\sigma$}_1\cdot\mbox{\boldmath $\sigma$}_2){\bf q}^2.
\label{eq:30.33} \end{equation}
The proper treatment of such a (non-local) Pauli-invariant 
has been developed for the ESC16-models, which is described in \cite{NRY19a}, Appendix~B .
For the treatment of the potentials with $P_8$ we use the identity \cite{BDI70}
\begin{equation}
 P_8 = -(1+\mbox{\boldmath $\sigma$}_1\cdot\mbox{\boldmath $\sigma$}_2) P_6.
\label{eq:30.34} \end{equation}
Under time-reversal $P_7 \rightarrow -P_7$ and $P_8 \rightarrow -P_8$.
Therefore for elastic scattering $V_7=V_8=0$. 
Anticipating the explicit results for the potentials in section~\ref{sec:3} we
notice the following:                     
 (i) For the general BB-reaction we will find no contribution to $V_7$. 
The operators $P_6$ and $P_8$ give spin singlet-triplet transitions. 
 (ii) In the case of non-strangeness-exchange ($\Delta S=0$), $V_6 \neq 0$  
and $V_8$=0. The latter follows from our approximation to neglect the
mass differences among the nucleons, between the $\Lambda$ and $\Sigma$'s, and
among the $\Xi$'s.
 (iii) In the case of strangeness-exchange ($\Delta S=\pm 1$), $V_6,V_8 \neq 0$.    
The contributions to $V_6$ come from graphs with both spin- and particle-exchange,
i.e. Majorana-type potentials having the $P_f P_\sigma P_6= -P_x P_6$-operator.
Here, $P_f P_\sigma$ reflect our convention for the two-particle wave functions,
see \cite{NRS77}. 
The contributions to $V_8$ come from graphs with particle-exchange and   
spin-exchange, because $P_8=-P_\sigma P_6$. Therefore, we only have to
apply $P_f$ in order to map the wave functions after such exchange onto
our two-particle wave-functions. So, we have the $P_f P_8= +P_x P_6$-operator.
Here, we used that for BB-systems the allowed physical states satisfy 
$P_f P_\sigma P_x=-1$.


\noindent In the SU(6) quark model \cite{LeY73}, 
instead of the Pauli-spinors, one uses for the quarks the Dirac-spinors 
\begin{equation}
 u^{(0)}_i({\bf p}_i) = \sqrt{\frac{E_i+m_i}{2 m_i}}
 \left[\begin{array}{c} 
 1 \\ 
\frac{\mbox{\boldmath \scriptsize{$\sigma$}}_i\cdot{\bf p}_i}{E_i+m_i}
 \end{array}\right] \otimes\ \chi_i,   
\label{eq:2.3} \end{equation}
where ${\bf p_i}_i$ denotes the three-momentum of the quarks in e.g. the CM-system.

\section{Extended-Soft-Core QQM and NNM Interactions}
\label{app:22}  
In the ESC-model there are single- and pair-meson quark-quark couplings.
They are the basis for the OBE, TME and MPE potentials. 
The nucleon-nucleon-meson (NNM) interactions have been described in 
detail in \cite{NRY19a}. Therefore, we restrict ourselves in this section 
to the quark-quark-meson (QQM) interactions.
The meson-quark couplings are designed such as to reproduce the ESC-potentials for
baryon-baryon when folded in with the constituent quark wave
functions of the SU(6) quark-model. Strictly, for the TME and MPE
potentials a modification should be made in the presence of
quark matter. In this paper such quark density corrections are omitted.

\subsection{Meson-quark-quark Interactions}
\label{app:22a}   
The potential of the ESC-model contains the contributions from 
(i) One-boson-exchanges, (ii) Uncorrelated 
Two-Pseudo-scalar exchange,  
and (iii) Meson-Pair-exchange. In this section we 
review the potentials and indicate the changes with respect to 
earlier papers on the OBE- and ESC-models.
The spin-1 meson-exchange is an important ingredient for the baryon-baryon force. 
In the ESC16-model we treat the vector-mesons and the axial-vector mesons 
according to the Proca- \cite{IZ80} and the B-field- \cite{Nak72,NO90} formalism
respectively. For details, we refer to \cite{NRY19a}, Appendix~C.

\subsection{One-Boson-Exchange Interactions in Momentum Space}
\label{app.3a}
The local interaction Hamilton densities for the different couplings
are \cite{BD65} \\ \\
        a) Pseudoscalar-meson exchange $(J^{PC}=0^{-+})$
         \begin{equation}
         {\cal H}_{pv}(x)= \frac{f_{pv}}{m_{\pi^{+}}}
         \bar{q}(x)\gamma_{\mu}\gamma_{5}
                q(x)\partial^{\mu}\phi_{P}(x). \label{eq:3.1a}\end{equation}
This is the pseudovector coupling, and the
relation with the pseudoscalar coupling is 
$g_p = 2m_Q/m_{\pi^+}$, where $m_Q$ is the quark mass.\\ \\     
        b) Vector-meson exchange $(J^{PC}=1^{--})$
       \begin{eqnarray}
   {\cal H}_v^{1)} &=& g_{v}\bar{q}(x)\gamma_{\mu}q(x)\phi_{V}^{\mu}
                +\frac{f_v}{4{\cal M}}\bar{q}(x)\sigma_{\mu\nu}
                q(x) (\partial^{\mu}\phi^{\nu}_{V}-\partial^{\nu}
                      \phi^{\mu}_{V}) \nonumber\\ &=& 
       \biggl[(\bar{q}(x)\gamma_\mu q(x))\ f_{1,v}+\frac{i}{2}\left(
       \bar{q}(x)\stackrel{\leftrightarrow}{\partial_\mu} q(x)\right)\ f_{2,v}
       \biggr]\cdot\phi_V^\mu, \nonumber\\
       \label{eq:3.2a}\end{eqnarray}
       where $\sigma_{\mu\nu}= i[\gamma_{\mu},\gamma_{\nu}]/2$,
       and $f_{1,v}=g_v+(m_Q/{\cal M}) f_v, f_{2,v}=-f_v/{\cal M}$.
       The scaling mass ${\cal M}$ will be taken to be the proton mass.         
 The Gordon decomposition 
 \begin{eqnarray*}
&& \partial_\nu\left[\bar{q}(x) \sigma^{\mu\nu} q(x)\right] = 2\bar{m}_Q
 \bar{q}(x) \gamma^\mu q(x) + i \bar{q}(x)\!\! \stackrel{\leftrightarrow}{\partial}^\mu\!\! q(x)
\label{eq:3.3a}\end{eqnarray*}
with $\bar{m}_Q=\left(m_Q^\prime + m_Q\right)$, 
shows that the magnetic-coupling consists of a pure vector and scalar bilinear quark-field
part. As deduced in \cite{TAR-QQ14}, an extra interaction is needed 
in order to give the correct structure of the baryon-baryon potential. 
Therefore, on the quark-level we add the interaction
\begin{equation}
   {\cal H}_v^{2)}=-\frac{\Box}{4m_Q^2} \biggl[ [\bar{q}(x)\gamma_\mu q(x)] f'_{1,v}
   +\left(i\bar{q}(x)\stackrel{\leftrightarrow}{\partial_\mu} q(x)\right) f'_{2,v}\biggr]
 \cdot\phi_{V}^{\mu}, \nonumber\\
\label{eq:3.4a}\end{equation}
where $f'_{1,v}=(4/9) f_{1,v}, f'_{2,v}= (4/9) f_{2,v}$.
Then, the total vector-exchange interaction is
\begin{eqnarray}
 {\cal H}_v &=& \bar{g}_{v}\bar{q}(x)\gamma_{\mu}q(x)\phi_{V}^{\mu}
             +\frac{\bar{f}_v}{4{\cal M}}\bar{q}(x)\sigma_{\mu\nu} q(x)
              (\partial^{\mu}\phi^{\nu}_{V}-\partial^{\nu} \phi^{\mu}_{V}), \nonumber\\
  \bar{g}_v &=& g_v\left(1-\frac{g'_v}{g_v}\frac{\Box}{4m_Q^2}\right),\ 
  \bar{f}_v = f_v\left(1-\frac{f'_v}{f_v}\frac{\Box}{4m_Q^2}\right). \nonumber\\
\label{eq:3.6a} \end{eqnarray}
An attractive alternative to the inclusion of the $(g_v',f_v')$-couplings would be to have a zero
in the QQV form factors. For $g'_v/g_v = f'_v/f_v=4/9$ this zero is at ${\bf k}^2=M_N^2$,
{\it i.e.} a short range effect.\\

\noindent c)\ Axial-vector-meson exchange ( $J^{PC}=1^{++}$, 1$^{st}$ kind):
\begin{equation}
 {\cal H}_a^{(1)} = g_a[\bar{q}(x)\gamma_\mu\gamma_5q(x)] \phi^\mu_A + \frac{if_a}{{\cal M}}
 [\bar{q}(x)\gamma_5q(x)]\ \partial_\mu\phi_A^\mu.
\label{eq:OBE.1}\end{equation}
We impose axial-current conservation by the relation 
$ f_a = \left(m_{A_1}^2/(2 m_Q{\cal M}\right)^{-1} g_a$ \cite{dercopax}.
The details of the treatment of the axial-vector mesons are given in 
\cite{NRY19a}, Appendix~B.
It was found in \cite{TAR-QQ14} that the correct reproduction of the baryon-baryon 
spin-orbit potential obtained by
a folding of the axial-exchange between quarks requires the additional interaction
\begin{equation}
 {\cal H}_a^{(2)} = -i \frac{g_a^\prime}{{\cal M}^2} \left\{\varepsilon_{\mu\nu\alpha\beta}
 [\partial^\alpha\bar{q}(x)\gamma^\nu \partial^\beta q(x)]\right\}\cdot\phi_A^\mu
\label{eq:OBE.2}\end{equation}
with $g_a^\prime = g_a$.
\\[0.2cm]
\noindent d)\ Axial-vector-meson exchange ( $J^{PC}=1^{+-}$, 2$^{nd}$ kind):
\begin{equation}
 {\cal H}_b = \frac{if_b}{m_B}
 [\bar{q}(x)\sigma_{\mu\nu}\gamma_5 q(x)\ \partial^\nu\phi_B^\mu\ .
\label{eq:OBE.3}\end{equation}
Like for the axial-vector mesons of the
1$^{st}$-kind we include an SU(3)-nonet with members $b_1(1235), h_1(1170), h_1(1380)$.
In the quark-model they are $Q\Bar{Q}(^1P_1)$-states.\\[0.2cm]
\noindent e)\ Scalar-meson exchange ($J^{PC}=0^{++}$):
\begin{equation}
 {\cal H}_s =  g_s\left\{ g_s -\frac{g_s^\prime}{g_s} \frac{\Box}{4m_Q^2}\right\}
[\bar{q}(x) q(x)]\cdot\phi_S, 
\label{eq:OBE.4}\end{equation}
with $g_s'/g_s = -8/9$. 
Again, the requirement from the folding of meson-exchange between quarks into
the baryon gives $g_s' \approx -g_s$. 
It is clear that inclusion of the $g_s'$ does not introduce a zero in the scalar-quark-quark
coupling. The additional contribution from the $g_s'$ coupling is taken onto
account easily.
In the ESC-models we include a zero in the form factor, which we also keep in the 
quark-quark potential.


\noindent f)\ Pomeron-exchange ($J^{PC}=0^{++}$):
       The vertices for this `diffractive'-exchange have the
       same Lorentz structure as those for scalar-meson-exchange.\\[0.2cm]
\noindent g)\ Odderon-exchange ($J^{PC}=1^{--}$):
\begin{equation}
 {\cal H}_O = g_O[\bar{\psi}\gamma_\mu\psi] \phi^\mu_O + \frac{f_O}{4{\cal M}}
 [\bar{\psi}\sigma_{\mu\nu}\psi] (\partial^\mu\phi^\nu_O-\partial^\nu\phi_O^\mu).
\label{eq:OBE.6}\end{equation}
Since the gluons are flavorless, Odderon-exchange is treated as an SU(3)-singlet.
Furthermore, since the Odderon represents a Regge-trajectory with an intercept
equal to that of the Pomeron, and is supposed not to contribute for small ${\bf k}^2$,
we include a factor ${\bf k}^2/{\cal M}^2$ in the coupling.\\[0.2cm]
 
Including form factors $f({\bf x}'-{\bf x})$ ,
the interaction hamiltonian densities are modified to
\begin{equation}
        H_{X}({\bf x})=\int\!d^{3}x'\,
  f({\bf x}'-{\bf x}){\cal H}_{X}({\bf x}'),
\end{equation}
 for $X= P,\ V,\ A$, and $S$ ($P =$ pseudo-scalar, $V =$ vector,
 $A=$ axial-vector, and $S =$ scalar). The         
potentials in momentum space are the same as for point interactions,
except that the coupling constants are multiplied by the Fourier
transform of the form factors.
 
In the derivation of the $V_{i}$ we employ the same approximations as in 
\cite{NRS78,MRS89}, i.e.
\begin{enumerate}
\item   We expand in $1/M$: 
    $E(p) = \left[ {\bf k}^{2}/4 +
    {\bf q}^{2}+M^{2}\right]^{\frac{1}{2}}$\\
    $\approx M+{\bf k}^{2}/8M + {\bf q}^{2}/2M$
 and keep only terms up to first order in ${\bf k}^{2}/M$ and
 ${\bf q}^{2}/M$. This except for the form factors where
 the full ${\bf k}^{2}$-dependence is kept throughout
 the calculations. Notice that the gaussian form factors
suppress the high ${\bf k}^{2}$-contributions strongly.
\item   In the meson propagators
$       (-(p_{1}-p_{3})^{2}+m^{2})  
        \approx({\bf k}^{2}+m^{2})$ .
\item   When two different baryons are involved at a $BBM$-vertex
        their average mass is used in the
        potentials and the non-zero component of the momentum transfer
        is accounted for by using an effective mass in
        the meson propagator (for details see \cite{MRS89}).     
\end{enumerate}
 
Due to the approximations we get only a linear dependence on
${\bf q}^{2}$ for $V_{1}$. In the following, separating the local and the
non-local parts, we write
\begin{equation}
  V_{i}({\bf k}^{2},{\bf q}^{2})=
  V_{i a}({\bf k}^{2})+V_{i b}({\bf k}^{2})({\bf q}^{2}+\frac{1}{4}{\bf k}^2),
\label{vcdec} \end{equation}
where in principle $i=1,8$. 
 
The OBE-potentials are now obtained in the standard way (see e.g.\
\cite{NRS78,MRS89}) by evaluating the $BB$-interaction in Born-approximation.
We write the potentials $V_{i}$ of Eqs.~(\ref{vcdec}) in the form
\begin{equation}
  V_{i}({\bf k}\,^{2},{\bf q}\,^{2})=
   \sum_{X} \Omega^{(X)}_{i}({\bf k}\,^{2})
   \cdot \Delta^{(X)} ({\bf k}^{2},m^{2},\Lambda^{2}).
\label{nrexpv2} \end{equation}
Furthermore for $X=P,V$ 
\begin{equation}
   \Delta^{(X)}({\bf k}^{2},m^{2},\Lambda^{2})= e^{-{\bf k}^{2}/\Lambda^{2}}/  
                    \left({\bf k}^{2}+m^{2}\right),
\label{propm1} \end{equation}
and for $X=S,A$ a zero in the form factor
\begin{equation}
   \Delta^{(S)}({\bf k}^{2},m^{2},\Lambda^{2})= \left(1-{\bf k}^2/U^2\right)\
  e^{-{\bf k}^{2}/\Lambda^{2}}/  
  \left({\bf k}^{2}+m^{2}\right),
\label{propm2} \end{equation}
and for $X=D,O$
\begin{equation}
   \Delta^{(D)}({\bf k}^{2},m^{2},\Lambda^{2})=\frac{1}{{\cal M}^{2}}
   e^{-{\bf k}^{2}/(4m_{P,O}^{2})}.
\label{Eq:difdel}
\end{equation}
In the latter expression ${\cal M}$ is a universal
scaling mass, which is again taken to be the proton mass.
The mass parameter $m_{P}$ controls the ${\bf k}^{2}$-dependence of
the Pomeron-, $f$-, $f'$-, $A_{2}$-, and $K^{\star\star}$-potentials.
Similarly, $m_O$ controls the ${\bf k}^2$-dependence of the Odderon.\\

\noindent {\it In the following we give the OBE-potentials in momentum-space for the 
hyperon-nucleon systems. From these those for NN and YY can be deduced easily.
We assign the particles 1 and 3 to be hyperons, and particles 2 and 4 to be
nucleons. Mass differences among the hyperons and among the nucleons will be neglected.}\\

\onecolumngrid

\begin{flushleft}
\rule{16cm}{0.5mm}
\end{flushleft}
\subsection{The Meson-Pair Interactions}
\label{sec:22b}
For the phenomenological SU(2) meson-pair interactions the Hamiltonians,
for meson-pairs with quantum numbers (J,P,C), for the non-strange quarks
 i.e. below $q(x) \equiv Q_1(x)$, are
\begin{subequations}
\label{HPs}
\begin{eqnarray}
  J^{PC}=0^{++}:\ {\cal H}_{S} &=& \bar{q}(x) q(x)\
       \left[g_{(\pi\pi)_{0}}\bm{\pi}\!\cdot\!\bm{\pi} +
             g_{(\sigma\sigma)}\sigma^{2}\right] / m_{\pi},  \label{LPs}\\
  {\cal H}_{E} &=& \bar{q}(x)\bm{\tau}q(x)\cdot\bm{\pi}
       \left[g_{(\pi\eta)}\eta + g_{(\pi\eta')}\eta'\right] / m_{\pi},
                                                             \label{LPe}\\
  {\cal H}_{S_2} &=& \bar{q}(x)q(x)\
       h_{(\pi\pi)_{0}}\partial_\mu\bm{\pi}\!\cdot\!\partial^\mu\bm{\pi} 
                                                 / m_{\pi}^3,  \label{LPs2}\\
  J^{PC}=1^{--}:\ {\cal H}_{V} &=& 
         g_{(\pi\pi)_{1}}\bar{q}(x)\gamma_{\mu}\bm{\tau}q(x)
         \cdot(\bm{\pi}\!\times\!\partial^{\mu}\bm{\pi})
         / m^{2}_{\pi}                                         \nonumber\\
  & &  -\frac{f_{(\pi\pi)_{1}}}{2{\cal M}}\bar{q}(x)\sigma_{\mu\nu}
         \bm{\tau}q(x)\partial^{\nu}\cdot(\bm{\pi}\!\times\!
         \partial^{\mu}\bm{\pi}) / m^{2}_{\pi},            \label{LPv}\\
  J^{PC}=1^{++}:\ {\cal H}_{A} &=& g_{(\pi\rho)_{1}}
         \bar{q}(x)\gamma_{5}\gamma_{\mu}\bm{\tau}q(x)\cdot
         (\bm{\pi}\times\bm{\rho}^{\mu}) / m_{\pi},      \label{LPa}\\
  {\cal H}_{P} &=& g_{(\pi\sigma)} \bar{q}(x)\gamma_{5}\gamma_{\mu}
         \bm{\tau}q(x) \cdot (\bm{\pi}\partial^{\mu}\sigma
         -\sigma\partial^{\mu}\bm{\pi}) / m^{2}_{\pi}      \nonumber\\
  & + & g_{(\pi P)} \bar{q}(x)\gamma_{5}\gamma_{\mu}
         \bm{\tau}q(x) \cdot (\bm{\pi}\partial^{\mu}P
         -P\partial^{\mu}\bm{\pi}) / m^{2}_{\pi},          \label{LPp}\\
  J^{PC}=1^{+-}:\ {\cal H}_{H} &=& -ig_{(\pi\rho)_{0}}
         \bar{q}(x)\gamma_{5}\sigma_{\mu\nu}q(x)\partial^{\nu}
         (\bm{\pi}\!\cdot\!\bm{\rho}^{\mu}) / m^{2}_{\pi},
                                                             \label{LPh}\\
  {\cal H}_{B} &=& -ig_{(\pi\omega)} \bar{q}(x)\gamma_{5}
         \sigma_{\mu\nu}\bm{\tau}q(x) \cdot \partial^{\nu}
         (\bm{\pi}\,\omega^{\mu}) / m^{2}_{\pi}.           \label{LPb}
\end{eqnarray}
\end{subequations}
For the SU(3) generalization see Ref.~\cite{Rij04b} section III.
 
In Eq.~(\ref{HPs}) also the Pomeron contribution is listed, but
in recent ESC-models $g_{(\pi P)}=0$.
The same is true for the ${\cal H}_{S_2}$ interaction, which we will discuss
in connection with the FM three-body force \cite{Miya56,Fuj57}.

\noindent {\it As for the scaling of the pair-coupling parameters, the
$\pi^+$-mass was choosen. For the operators $\partial^\mu\pi(x)$ this follows the 
non-linear chiral models. The other scaling $m_\pi$-factors may be could be
better replaced by $M$, the nucleon mass. This would presumably represent
better the scale of the physics involved. For example pair-couplings from
$N\bar{N}$-pairs ('negative-energy states') would be parameterized more
naturally this way. However, in our works  on the ESC-model we sofar always
used the $m_\pi$-mass as a scaling parameter, and therefore we will do this
also in this paper}.\\


 

\begin{flushleft}
\rule{16cm}{0.5mm}
\end{flushleft}
 
\section{Channels, Potentials, and $SU(3)$ Symmetry}
\label{sec:12}

\subsection{Channels and Potentials}     
\label{sec:2a}
In this paper we consider the quark-quark and quark-nucleon
reactions with strangeness S=0: 
\begin{subequations}\label{eq:2.1}
\begin{eqnarray}
	Q(s_a,i_a)+Q(s_b,i_b) &\rightarrow& Q(s'_a,i'_a)+Q(s'_b,i'_b), \\         
	Q(s_a,i_a)+N(s_b,i_b) &\rightarrow& Q(s'_a,i'_a)+N(s'_b,i'_b),          
\end{eqnarray}\end{subequations}
where $Q=U,D$ and $N=P,N$ are iso-doublets.
The  3-components of the spins and isospins are $\pm 1/2$.
Like in Ref.'s~\cite{MRS89} we will also refer to $a$ and $a'$
as particles 1 and 3, and to $b$ and $b'$ as particles 2 and 4. For the 
kinematics and the definition of the amplitudes, we refer to papers  
\cite{Rij04a,Rij04b} of the series of papers on the ESC04 model. 
Here we note that both the BB- and QQ-channels are of the same type, 
nanely spin-1/2-spin 1/2 scattering.
Similar material can be found in \cite{MRS89}.
Also, in paper I the derivation of the Lippmann-Schwinger equation 
in the context of the relativistic two-body equation is described.

For the quark-quark and quark-nucleon there are three channels with different 
 charges for the baryon-number B:
\begin{subequations} \label{eq:2.2}
\begin{eqnarray}
 B=2/3:\ \left\{\begin{array}{ll}
 q=+4/3:\ \  &  UU \rightarrow\ UU, \\               
 q=+1/3:\ \  &  UD \rightarrow\ UD,  \\               
 q=-2/3:\ \  &  DD \rightarrow\ DD,  \end{array}\right. \\
 & & \nonumber\\
 B=4/3:\ \left\{\begin{array}{ll}
 q=+5/3:\ \  &  UP \rightarrow\ UP, \\               
 q=+2/3:\ \  &  UN \rightarrow\ UN,  \\               
 q=+2/3:\ \  &  DP \rightarrow\ DP,  \\               
 q=-1/3:\ \  &  DN \rightarrow\ DN,  \end{array}\right. \\
\end{eqnarray}
\end{subequations}
{\it Note that since the spin and isospin of these quarks and the nucleons
are identical the spin and isospin factors of the potentials are 
completely analogous to that for the nucleon-nucleon potentials.
Therefore, we refer for these factors to the NN-papers, e.g. \cite{Rij04a}
etc..}

\noindent Like in \cite{MRS89,Rij04a}, the potentials are calculated on the isospin basis.
there are only two isospin channels:
(i) $ I= 1:\ \ (UU, (UD+DU)/\sqrt{2}, DD)$, and 
(ii) $ I=0:\ \ (UD-DU)/\sqrt{2}$. Similarly for the QN-channels.


In this work we give the QQ- and the QN-potentials for the Lippmann-Schwinger 
equation in momentum space, and
the Schr\"{o}dinger equation in configuration space.             

The momentum space and configuration space potentials for the ESC models
have been described in papers \cite{Rij04a} and \cite{NRY19a} for baryon-baryon in general.
Also in the ESC-model, the potentials are of such a form that they are exactly 
equivalent in both momentum space and configuration space. 
The treatment of the mass differences among the quarks are handled exactly 
similar as is done in \cite{MRS89}. Also, exchange potentials related to
strange meson exchange $K, K^*$ etc. , can be found in these references. 

The quark mass differences in the intermediate states for TME- and MPE-
potentials will be neglected for QQ-scattering. This, although possible
in principle, becomes rather laborious and is not expected to change the 
characteristics of the quark-quark potentials much.

\subsection{QQM-couplings in $SU(3)$, Matrix-representation}
\label{sec:12b}
The $Q=(U,D,S)$-quarks are in the fundamental $\{3\}$-irrep, and in matrix
notation represented by a collumn.
In previous work of the Nijmegen group, e.g. \cite{MRS89},
the treatment of $SU(3)$ has been given in detail for the BBM interaction
Lagrangians and the coupling coefficients of the OBE-graphs. However,
for the ESC-models we also need the coupling coefficients for the TME- and the
MPE-graphs. Since there are many more TME- and MPE-graphs than OBE-graphs, an
computerized computation is desirable. As in the baryon-baryon papers,
here the so-called 'cartesian-octet'-representation for the mesons 
is quite useful. 
Therefore, we give an exposition of this 
representation, its connection with the matrix representation used in our previous
work, and the formulation of the coupling coefficients used in the 
automatic computation. 

The various meson nonets (we take the pseudoscalar mesons with $J^P=0^+$
as an example), see e.g. \cite{Swa63,Car66}, are represented by 
\begin{equation}
     P=P_{\{1\}}+P_{\{8\}},
\label{eq:2.4}\end{equation}
where the singlet matrix $P_{\{1\}}$ has elements $\eta_0/\sqrt{3}$
on the diagonal, and the octet matrix $P_{\{8\}}$ is given by
\begin{equation}
   P_{\{8\} } = \left( \begin{array}{ccc}
      {\displaystyle\frac{\pi^{0}}{\sqrt{2}}+\frac{\eta_{8}}{\sqrt{6}}}
             & \pi^{+}  &  K^{+}  \\[2mm]
      \pi^{-} & {\displaystyle-\frac{\pi^{0}}{\sqrt{2}}
         +\frac{\eta_{8}}{\sqrt{6}}}  &   K^{0} \\[2mm]
      K^{-}  &  \overline{{K}^{0}}
             &  {\displaystyle-\frac{2\eta_{8}}{\sqrt{6}}}
             \end{array} \right).
\label{eq:2.5}\end{equation}

The $SU(3)$-invariant BBP-interaction
Lagrangian can be written as \cite{Swa63}
\begin{eqnarray}
   {\cal H}_{I} &=& g_{8}
 \sum_{p=1}^8\left[\bar{Q}_a\left(\lambda_p\right)_{ab} Q_b\right] \phi_{8,p}
 + g_1\ \left[\bar{Q} Q\right]\ \phi_9.
\label{eq:2.6}\end{eqnarray}
where $g_8$ and $g_1$ are the singlet and octet couplings. 
We write the octet coupling in the form of the meson matrix M:
\begin{equation}
   {\cal H}_I{(8)} = g_{8}\sqrt{2}\left[\bar{Q} M^{(8)} Q_b\right],\ \     
 M^{(8)}_{ab} = \sum_{p=1}^8\left(\lambda_p\right)_{ab} \phi_{8,p}.
\label{eq:2.7}\end{equation}

The convention used for the isospin doublets is
\begin{eqnarray}
&&  n=\left(\begin{array}{c} u \\ d \end{array} \right), \ 
  K=\left(\begin{array}{c} K^{+} \\ K^{0} \end{array} \right),
  \ K_{c}=\left(\begin{array}{c} \overline{K^{0}} \\
               -K^{-} \end{array} \right).        
\label{eq:2.8}\end{eqnarray}
Working out (\ref{eq:2.6}) on the isospin basis we have
\onecolumngrid
\begin{eqnarray}
 {\cal H}_I(8) &=& g_8\sqrt{2}\left(\bar{u},\bar{d},\bar{s}\right)
 \left(\begin{array}{ccc}
 \frac{\pi^0}{\sqrt{2}}+\frac{\eta_8}{\sqrt{6}} & \pi^+ & K^+ \\
  \pi^- &\frac{\pi^0}{\sqrt{2}}+\frac{\eta_8}{\sqrt{6}} & K^0 \\
  K^- & \bar{K}^0 & -\frac{2\eta_8}{\sqrt{6}} \end{array}\right)\
 \left(\begin{array}{c} u \\ d \\ s \end{array}\right) 
\nonumber\\ 
&=& g_8 \left[\vphantom{\frac{A}{A}}
 \bar{n}(\bm{\tau}\cdot\bm{\pi}) n + \sqrt{2}\left((\bar{n}\cdot K) s 
 +\bar{s} (\Bar{K}\cdot n)\right) + \frac{1}{\sqrt{3}} (\bar{n}\ n) \eta_8
 -\frac{2}{\sqrt{3}} (\bar{s} s) \eta_8 \right]
 \nonumber\\ &=& 
  g_{nn\pi}\ \bar{n}(\bm{\tau}\cdot\bm{\pi})n + g_{snK}\left(
  (\bar{n}\cdot K) s + \bar{s}(\bar{K}\cdot n)\right)
  +g_{nn\eta} (\bar{n} n) \eta_8 + g_{ss \eta} (\bar{s} s)\ \eta_8.
\label{eq:2.9}\end{eqnarray}
Here, we introduced the isospin-basis couplings
\begin{eqnarray}
&&  g_{nn\pi} = g_8\ ,\ g_{snk} = \sqrt{2} g_8\ ,\          
 g_{nn\eta} = \frac{1}{\sqrt{3}} g_8\ ,\ g_{ss\eta} = -\frac{2}{\sqrt{3}} g_8.
\label{eq:2.10}\end{eqnarray}
These couplings are similar to the OBE-couplings in baryon-baryon, 
and convenient for the transcription of the OBE-potentials from 
baryon-baryon to quark-quark. 

\noindent The precise connection with the couplings of ESC models is
given in Appendix~\ref{app:qpc}, where the $(g_8,g_1)$ are defined in
the framework of the quark-pair-creation (QPC) model. Furthermore, the
connection between QQM-couplings in the constituent quark-model (CQM) and 
the BBM-couplings gives a direct determination of the QQM-couplings
from the NN and YN data fitting.


\subsection{ Computations for OBE-, TME-, MPE-graphs SU(3)-factors}
\label{sec:12x}
The evaluation of the SU(3)-factors for the different potential graphs runs for
QQ/QN and NN/YN very similar. They can readily be copied from the calculation for
NN and YN given in Ref.~\cite{Rij04a,Rij04b}, where a Cartesian presentation \cite{Car66}
is exploited forthe computation of the SU(3) factors.

\subsection{Form Factors}                                                      
\label{sec:4b}
Also in this work, like in the NSC97-models \cite{RSY99}, the form 
factors depend on the SU$(3)$ assignment of the mesons,  
In principle, we introduce form factor masses 
$\Lambda_{8}$ and $\Lambda_{1}$ for the $\{8\}$ and $\{1\}$ members of each 
meson nonet, respectively. In the application to $YN$ and $YY$, we allow
for SU$(3)$-breaking, by using different cut-offs for the strange mesons
$K$, $K^{*}$, and $\kappa$. Moreover, for the $I=0$-mesons we assign the 
cut-offs as if there were no meson-mixing. For example we assign $\Lambda_1$ 
for $\eta', \omega, \epsilon$, and $\Lambda_8$ for $\eta, \phi, S^*$, etc.
\section{ Relation QQM- and  BBM-couplings}                   
\label{sec:5} 
In \cite{TAR-QQ14} the relation between the QQM- and BBM-couplings is
determined by requiring that the 1/M-expansion of the baryon-baryon
potentials is reproduced by folding, using the SU(6) quark-model
\cite{LeY73}. The relations are\\

\noindent (a) Pseudoscalar mesons: 
\begin{equation}
 f^{p}_{QQ\pi} =  f^{P}_{BB\pi},
\label{eq:5.1}\end{equation}
and similar relations for the $\eta, K, \eta^\prime$.
This follows from $g^p_{QQ\pi} = g^p_{BB\pi}/3$ and $m_q = M_B/3$. \\[2mm]
\noindent (b) Vector mesons: 
\begin{equation}
 g^{v}_{QQ\rho} = \frac{1}{3} g^V_{BB\rho}\ ,\ 
 f^{v}_{QQ\rho} = \frac{1}{3} f^V_{BB\rho},
\label{eq:5.2}\end{equation}
and similar relations for $\phi, K^*,\omega$.\\[2mm]
\noindent (c) Scalar mesons: 
\begin{equation}
 g^{s}_{QQa_0} = \frac{1}{3} g^S_{BBa_0}, 
\label{eq:5.3}\end{equation}
and similar relations for $f_0(993),\kappa, \epsilon=f_0(620)$.\\[2mm]
\noindent (d)  Axial-vector mesons (I): 
\begin{equation}
 g^{a}_{QQA_1} = \frac{1}{3} g^A_{BBA_1}\ ,\ 
 f^{a}_{QQA_1} = \frac{1}{3} f^A_{BBA_1},
\label{eq:5.4}\end{equation}
and similar relations for $D_1(1285)$, $K_A(1336)$, $E_1(1420)$.\\[2mm]
\noindent (e) Axial-vector mesons (II): 
\begin{equation}
 f^{b}_{QQB_1} = \frac{1}{3} f^B_{BBB_1}, 
\label{eq:5.5}\end{equation}
 and similar relations for $D_1(1285)$, $K_B(1300)$, and $E_1(1420)$.\\[2mm]
\noindent (f) Diffractive exchanges: Under the usual assumption of the
quark-additivity of the pomeron couplings \cite{Greenberg64} one has 
$g_{QQP} = g_{NNP}/3$, and similarly for the odderon couplings.


\onecolumngrid
\section{ Gluon and Confining Potentials}           
\label{sec:conf}       
The one gluon-exchange (OGE) has the form
\begin{equation}
 V_{OGE} = A\ (\bm{\lambda}_1\cdot\bm{\lambda}_2)\ V_V(m_G,r,\Lambda_G),
\label{sec:conf1}\end{equation}
where $V_V$ is the OBE vector exchange potential. Here, $m_G = 480$ MeV,
which is the mass of the gluon propagator in the
"liquid instanton model" \cite{Hut95}.
\noindent In \cite{RGGD12,Rib80}
the confining potential is taken to be a scalar
color-octet exchange potential. In \cite{Gloz97} the confining potential 
is color-singlet scalar exchange of the form
\begin{equation}
 V_{conf} = C_0 + C_1\ (\bm{\lambda}_1\cdot\bm{\lambda}_2)\ r^2,
\label{sec:conf2}\end{equation}
where $C_0$ is adjusted to give the 939 MeV for the nucleon mass, and
depends on the other parts of the total Q-Q potential. For the GBE-model
\cite{Gloz96a,Gloz96b} in \cite{Gloz97} table III the fitted GBE parameters are
$C_0$= -416 MeV, $C_1$= 2.33.\\
\noindent Since the GBE-model approach is also that of Manohar-Georgi, we
choose in this work the confining potential in (\ref{sec:conf2}).

\begin{table}
 \caption{Color and Spin matrix elements, ${\bf F}=\bm{\lambda}/2$.}
\begin{center}
\begin{ruledtabular}
\begin{tabular}{|ccc|cc} & & & & \\
  S & I & C & $\langle \bm{\lambda}_1\cdot\bm{\lambda}_2\rangle$ &
 $\langle \bm{\sigma}_1\cdot\bm{\sigma}_2\rangle$ \\
 & & & & \\ \hline
 0 & 0 & $\{3^*\}$ &-8/3 & -3 \\
 0 & 1 & $\{6\}$   &+4/3 & -3 \\
 1 & 0 & $\{6\}$   &+4/3 & +1 \\
 1 & 1 & $\{3^*\}$ &-8/3 & +1 \\
\end{tabular}
\end{ruledtabular}
\end{center}
\label{tab:ff1}
\end{table}

\section{SU(3) NJL-form Instanton Potentials}                          
\label{sec:njl3}      
For SU(2) with $\psi=(u,d)$ and $\tau_0={\bf 1}$, the 't Hooft quark-quark
interaction reads
\begin{equation}
{\cal L}_{ud} = G_I\left[\vphantom{\frac{A}{A}} 
 (\bar{\psi}\tau_0\psi)^2+(\bar{\psi}i\gamma_5\bm{\tau}\psi)^2
 -(\bar{\psi}\bm{\tau}\psi)^2-(\bar{\psi}i\gamma_5\tau_0\psi)^2\right],
\label{njl3.0}\end{equation}
The SU(3) generalization of the 't Hooft interaction for the (u,d,s) quarks 
in the NJL-form reads
\begin{equation}
{\cal L}_{uds} = G_I\left[\vphantom{\frac{A}{A}} 
 (\bar{\psi}\lambda_0\psi)^2+(\bar{\psi}i\gamma_5\bm{\lambda}\psi)^2
 -(\bar{\psi}\bm{\lambda}\psi)^2-(\bar{\psi}i\gamma_5\lambda_0\psi)^2\right],
\label{njl3.1}\end{equation}
with $G_I=\lambda_{ud}/4$, and 
where $\psi =(u,d,s)$ i.e. the flavor $\{3\}$-irrep spinor field, 
$\lambda_0= \sqrt{4/3}\ {\bf 1}$, and $\lambda_a, a=1,8$ are the Gell-Mann matrices.\\

\noindent {\bf 1.\ Diagonal Potentials}: Working out the diagonal terms we have
\begin{eqnarray*}
{\cal L}_{uds} \Rightarrow  G_I 
\bigl(\lambda_{0,1}\lambda_{0,2}-\bm{\lambda}_1\cdot\bm{\lambda}_2\bigr)
 \left[\vphantom{\frac{A}{A}} 
 (\bar{q}_i q_i)^2+(\bar{q}_i\gamma_5 q_i)^2\right],
\end{eqnarray*}
with i=u,d,s.
In the CM-system assigning the momenta $({\bf p}, -{\bf p})$ in the initial
state and $({\bf p}',-{\bf p}')$ in the final state one has
\begin{eqnarray*}
 (\bar{q} q)^2 &\rightarrow& 1-\frac{1}{4M^2}\bigl[2{\bf p}'\cdot{\bf p}
 +i(\bm{\sigma}_1+\bm{\sigma}_2)\cdot{\bf p}'\times{\bf p}\bigr], \\
 (\bar{q}\gamma_5 q)^2 &\rightarrow&-\frac{1}{4M^2}\
 \bm{\sigma}_1\cdot({\bf p}'-{\bf p})\ \bm{\sigma}_2\cdot({\bf p}'-{\bf p})
\end{eqnarray*}
Using the variables ${\bf k}={\bf p}'-{\bf p}$ and ${\bf q}=({\bf p}'+{\bf p})/2$ 
the potential becomes



\begin{eqnarray} 
&& \widetilde{V}({\bf p}',{\bf p}) = -2 G_I\ 
\bigl(\lambda_{0,1}\lambda_{0,2}-\bm{\lambda}_1\cdot\bm{\lambda}_2\bigr)
\biggl[ 1+\left(1-\frac{1}{3}\bm{\sigma}_1\cdot\bm{\sigma}_2\right)\ \frac{{\bf k}^2}{4M^2}
 -\frac{{\bf q}^2+{\bf k}^2/4}{2M^2} 
\nonumber\\ && 
 -\frac{1}{4M^2}\left(   
\bm{\sigma}_1\cdot{\bf k} \bm{\sigma}_2\cdot{\bf k} -\frac{1}{3}
 \bm{\sigma}_1\cdot\bm{\sigma}_2\ {\bf k}^2\right) 
 +\frac{i}{4M^2} (\bm{\sigma}_1+\bm{\sigma}_2)\cdot{\bf n} 
+ \frac{1}{16M^4} (\bm{\sigma}_1\cdot{\bf n})(\bm{\sigma}_2\cdot{\bf n})
\biggr], 
\nonumber\\
\label{njl3.3}\end{eqnarray}
where ${\bf n} = {\bf q}\times{\bf k}$, and the quadratic-spin-orbit term is added for completenes.\\
Adding a gaussian cut-off $F_I({\bf k}^2)= \exp\bigl[-{\bf k}^2/(\Lambda^2\bigr]$,
with $m_I=\Lambda/2$, the local instanton potentials become, apart from the
flavor factor,
\begin{eqnarray}
 V_I &=& -\frac{2g_I}{\pi\sqrt{\pi}}\frac{m_I^3}{\Lambda^2} \biggl[
 1+\frac{m_I^2}{2M^2}\left(3-2m_I^2r^2\right)
 \left(1-\frac{1}{3}\bm{\sigma}_1\cdot\bm{\sigma}_2\right)
 +\frac{m_I^2}{3M^2}\ (m_Ir)^2\ S_{12} \nonumber\\ && 
 +\frac{m_I^2}{M^2}\ {\bf L}\cdot{\bf S}
 +\frac{m_I^4}{M^4}\ Q_{12}
\biggr]\ \exp\bigl[-m_I^2r^2\bigr].
\label{njl3.4}\end{eqnarray}

\noindent Taking $\Lambda = 1$ GeV/c$^2$, $G_I=\lambda_{ud}/4$ the coupling   
$g_I = G_I\Lambda^2= 2.0-2.5$. 


\twocolumngrid

For u,d quarks the flavor factor becomes, see also (\ref{eq:inst1}),
 $(1-\bm{\tau}_1\cdot\bm{\tau}_2)$,
which gives 0 and 4 for I=1 and I=0 respectively. 
For $m_I=200$ MeV and $M=m_Q=M_N/3 \approx 315$ MeV, the factor 
$(1+3m_I^2/2M^2) \approx 1.6$.
This gives $V_I(^1S_0,I=1)=0$ and $V_I(^3S_1,I=0) < 0$ for r=0.\\
In analyzing the U(1)-problem, Weinberg \cite{Weinberg75} chooses 
$\lambda_0=\sqrt{2/3}\ {\bf 1}$ giving for u,d quarks $(1/3-\bm{\tau}_1\cdot\bm{\tau}_2)$
which is -2/3 and 10/3 for I=1 and I=0 respectively. This gives repulsion and
attraction for respectively $^1S_0(I=1)$ and $^3S_1(I=0)$.

\noindent The non-local term in (\ref{njl3.3}) has the same sign as for scalar 
and vector exchange, and opposite to Pomeron exchange. 
Therefore, compare \cite{NRS78} formula (34), one has
\begin{eqnarray}
 V_{n.l.}(r) &=& - G_I\ \biggl\{\bm{\nabla}^2 
 \exp\biggl[-\frac{1}{4}\Lambda^2r^2\biggr] + 
 \exp\biggl[-\frac{1}{4}\Lambda^2r^2\biggr] \bm{\nabla}^2\biggr\} \nonumber\\
 &\equiv& -\bigg[\bm{\nabla}^2\frac{\phi(r)}{2M_{red}}+\frac{\phi(r)}{2M_{red}}
 \biggr], 
\label{njl3.6}\end{eqnarray}
which, with $M_{red}=M/2$, gives
\begin{eqnarray}
 \phi(r) &=& +(G_IM^2)\left(\frac{\Lambda}{2\sqrt{\pi}M}\right)^3\
 \exp\left[-\frac{1}{4}\Lambda^2r^2\right].
\label{njl3.7}\end{eqnarray}
Now, $(\Lambda/2\sqrt{\pi}M) \approx 0.85$ and 0.56 for the u,d and s quark
respectively. For $g_I=G_IM^2 = 2.0-2.5$ the non-local function $\phi(r)$ is
not small.\\
\noindent The flavor factor for the non-strange quarks becomes
$(1-\bm{\tau}_1\cdot\bm{\tau}_2)$, which is due to the choice for $\lambda_0$.\\

\noindent {\bf 2.1.\  Non-diagonal Potentials}: There are no non-diagonal terms!?     For example $s \rightarrow u$: 
\begin{eqnarray*}
 (\bar{\psi}\bm{\lambda}\psi)^2 \rightarrow 
 (\bar{\psi}\lambda_4\psi)^2 + (\bar{\psi}\lambda_5\psi)^2 \rightarrow
 (\bar{u}s)^2-(\bar{u}s)^2 =0, etc.
\end{eqnarray*}

\section{ ESC16-model: Fitting $NN\oplus YN \oplus YY$-data} 
\label{sec:15a} 
In the simultaneous $\chi^2$-fit of the $NN$-, $YN$-, and YY-data a 
{\it single set of parameters} was used, which means the same parameters for all 
BB-channels.
The input $NN$-data are the same as in Ref.~\cite{Rij04a}, and we refer the reader 
to this paper for a description of the employed phase shift analysis 
\cite{Sto93,Klo93}. 

It appeared that the OBE-couplings could be constrained successfully
by the 'naive' predictions of the QPC-model \cite{Mic69,LeY73}. Although these 
predictions, see section \ref{sec:4}, are 'bare' ones, the policy was to keep 
the many OBE-couplings in the neighborhood of the QPC-values.
Also, it appeared that we could either fix the $F/(F+D)$-ratios 
to those as suggested by the QPC-model, 
or apply the same restraining strategy as for the OBE-couplings.                      
\subsection{ Fitted BB-parameters}                       
\label{sec:15b} 
The treatment of the broad mesons $\rho$ and $\epsilon$ was similar to that in the 
OBE-models \cite{NRS78,MRS89}. For the $\rho$-meson the same parameters are used 
as in these references. However, for the $\epsilon=f_0(620)$ assuming 
$m_\epsilon=620$ MeV and $\Gamma_\epsilon = 464$ MeV the Bryan-Gersten parameters      
\cite{Bry72} are used. For the chosen mass and width they are: 
$ m_1=496.39796$ MeV, $m_2=1365.59411$ MeV, and $\beta_1=0.21781, \beta_2=0.78219$.
Other meson masses are given in Table~\ref{table4}.
The sensitivity for the values of the cut-off masses of the $\eta$ and $\eta'$ 
is very weak. 
Therefore we have set the \{1\}-cut-off imass for the pseudoscalar nonet equal to 
that for the \{8\}.  Likewise, for the two nonets of the axial-vector mesons, 
see table~\ref{table5}.

Summarizing the parameters for baryon-baryon (BB) are:\\
 (i) NN Meson-couplings: $f_{NN\pi},f_{NN\eta'}$,     
 $ g_{NN\rho}, g_{NN\omega}$, 
  $f_{NN\rho},f_{NN\omega}$, $g_{NNa_0},g_{NN\epsilon}$,
   $g_{NNa_1}$, $f_{NNa_1}$, $g_{NNf'_1}$, $f_{NNf'_1}$, 
  $f_{NNb_1}$, $f_{NNh'_1}$\\
  (ii) $F/(F+D)$-ratios: $\alpha^{m}_{V}$, $\alpha_{A}$ \\
  (iii) NN Pair couplings: $g_{NN(\pi\pi)_1},f_{NN(\pi\pi)_1}$, $g_{NN(\pi\rho)_1}$,
  $g_{NN\pi\omega}, g_{NN\pi\eta}, g_{NN\pi\epsilon}$ \\
  (iv) Diffractive couplings and masslike parameters $g_{NNP}$, $g_{NNO}$, $f_{NNO}$, $m_P$, $m_O$ \\
  (v) Cut-off masses: $\Lambda_{8}^P = \Lambda_{1}^P$, $\Lambda_{8}^V$, $\Lambda_{1}^V$,
  $\Lambda_{8}^S$, $\Lambda_{1}^S$, and  $\Lambda_{8}^A$ = $\Lambda_{1}^A$.
 
The pair coupling $g_{NN(\pi\pi)_0}$ was kept fixed at zero.    
Note that in the interaction Hamiltonians of the pair-couplings (\ref{HPs}) 
the partial derivatives are scaled by $m_\pi$, and there is a scaling mass $M_N$.

The ESC models, are fully consistent with $SU(3)$-symmetry  
using a straightforward extension of the NN-model to YN and YY. 
This is the case for the OBE- and
TPS-potentials, as well as for the Pair-potentials.
All $F/(F+D)$-ratio's are taken as fixed with heavy-meson saturation in mind.

\subsection{ Coupling Constants, $F/(F+D)$ Ratios, and Mixing Angles}             
\label{sec:15c} 
In Table~\ref{table5} we give the ESC16 meson masses, and the fitted 
couplings and cut-off parameters \cite{NRY19a,NRY19b}. 
Note that the axial-vector couplings for the
B-mesons are scaled with $m_{B_1}$.
The mixing for the pseudo-scalar, vector, and scalar mesons, as well as the 
handling of the diffractive potentials, has been described elsewhere, see
e.g. Refs.~\cite{MRS89,RSY99}. The mixing scheme of the axial-vector mesons is completely
similar as for the vector etc. mesons, except for the mixing angle.        
As mentioned above, we searched for solutions where 
all OBE-couplings are compatible with the QPC-predictions. This time the QPC-model
contains a mixture of the $^3P_0$ and $^3S_1$ mechanism, whereas in 
Ref.~\cite{Rij04a} only the $^3P_0$-mechanism was considered.
For the pair-couplings all $F/(F+D)$-ratios were fixed to the predictions of
the QPC-model. 

\begin{table}
\caption{Meson couplings and parameters employed in the ESC16-potentials.
         Coupling constants are at ${\bf k}^{2}=0$.
         An asterisk denotes that the coupling constant is constrained via SU(3).
         The masses and $\Lambda$'s are given in MeV.}
\label{table4}
\begin{center}
\begin{ruledtabular}
\begin{tabular}{crccr} \hline\hline
meson & mass & $g/\sqrt{4\pi}$ & $f/\sqrt{4\pi}$ & \multicolumn{1}{c}{$\Lambda$}\\
\hline
 $\pi$         &  138.04 &           &   0.2684   &  1030.96\     \\
 $\eta$        &  547.45 &           & \hspace{1mm}0.1368$^\ast$   & ,, \hspace{5mm} \\
 $\eta'$       &  957.75 &           &   0.3181   &  ,, \hspace{5mm} \\ \hline
 $\rho$        &  768.10 &  0.5793   &   3.7791   &    680.79\    \\
 $\phi$        & 1019.41 &--1.2384$^\ast$ & \hspace{2mm}2.8878$^\ast$ & 
 ,, \hspace{5mm}  \\
 $\omega$      &  781.95 &  3.1149   & --0.5710 \hspace{0.5mm}   &    734.21\\ \hline
 $a_1 $        & 1270.00 &--0.8172   & --1.6521   &   1034.13\   \\
 $f_1 $        & 1420.00 &  0.5147 &   4.4754  &    
  ,,  \hspace{5mm} \\
 $f_1'$        & 1285.00 &--0.7596   & --4.4179 &      ,, \hspace{5mm}  \\ \hline
 $b_1 $        & 1235.00 &           & --2.2598   &   1030.96    \\
 $h_1 $        & 1380.00 &           & \hspace{2mm}--0.0830$^\ast$   &     
 ,,  \hspace{5mm} \\
 $h_1'$        & 1170.00 &           & --1.2386   &      ,, \hspace{5mm}  \\ \hline
 $a_{0}$       &  962.00 &  0.5393   &            &    830.42\    \\
 $f_{0}$       &  993.00 &--1.5766$^\ast$   &            &  
 ,, \hspace{5mm}  \\
 $\varepsilon$ &  620.00 &  2.9773   &            &   1220.28 \\ \hline
 Pomeron       &  212.06 &  2.7191   &            &              \\
 Odderon       &  268.81 &  4.1637   & --3.8859   &              \\
\hline
\end{tabular}
\end{ruledtabular}
\end{center}
\label{table5}
\end{table}
One notices that all the BBM $\alpha$'s have values rather close to that 
which are expected from the QPC-model. In the ESC16 solution $\alpha_A \approx 0.38$,
which is close to $\alpha_A \sim 0.4$. 
As in previous works, e.g. Ref.~\cite{NRS78}, $\alpha_V^e=1$ is 
kept fixed.
Above, we remarked that the axial-nonet parameters may be sensitive to whether
or not the heavy pseudoscalar nonet with the $\pi$(1300) are included.
In Table~\ref{table4} we show the OBE-coupling constants and the 
gaussian cut-off's $\Lambda$. The used  $\alpha =: F/(F+D)$-ratio's 
for the OBE-couplings are:
pseudo-scalar mesons $\alpha_{pv}=0.365$, 
vector mesons $\alpha_V^e=1.0, \alpha_V^m=0.472$, 
and scalar-mesons $\alpha_S=1.00$, which is calculated using the physical 
$S^* =: f_0(993)$ coupling etc.
\begin{table}[hbt]
\caption{Pair-meson coupling constants employed in the ESC16 MPE-potentials.     
         Coupling constants are at ${\bf k}^{2}=0$.
         The F/(F+D)-ratio are QPC-predictions, except that 
 $\alpha_{(\pi\omega)}=\alpha_{P}$, which is very close to QPC.}
\label{tab.gpair}
\begin{center}
\begin{ruledtabular}
\begin{tabular}{cclrc} \hline\hline
 $J^{PC}$ & $SU(3)$-irrep & $(\alpha\beta)$  &\multicolumn{1}{c}{$g/4\pi$} & $F/(F+D)$ \\
 \hline \\
 $0^{++}$ & $\{1\}$  & $g(\pi\pi)_{0}$   &  ---    &  ---    \\
 $0^{++}$ & ,,       & $g(\sigma\sigma)$ &  ---    &  ---    \\
 $0^{++}$ &$\{8\}_s$ & $g(\pi\eta)$      & -0.6894 &  1.000  \\ \hline
 $1^{--}$ &$\{8\}_a$ & $g(\pi\pi)_{1}$   &  0.2519 &  1.000  \\
          &          & $f(\pi\pi)_{1}$   &--1.7762 &  0.400  \\ \hline
 $1^{++}$ & ,,       & $g(\pi\rho)_{1}$  &  5.7017 &  0.400  \\
 $1^{++}$ & ,,       & $g(\pi\sigma)$    &--0.3899 &  0.400  \\
 $1^{++}$ & ,,       & $g(\pi P)$        &  ---    &  ---    \\ \hline
 $1^{+-}$ &$\{8\}_s$ & $g(\pi\omega)$    &--0.3287 &  0.365  \\
\hline
\end{tabular}
\end{ruledtabular}
\end{center}
\end{table}
In Table~\ref{tab.gpair} we list the fitted Pair-couplings for the MPE-potentials.
We recall that only One-pair graphs are included, in order to avoid double
counting, see Ref.~\cite{Rij04a}. The $F/(F+D)$-ratios are all fixed, assuming heavy-boson 
domination of the pair-vertices. The ratios are taken from the QPC-model for 
$Q\bar{Q}$-systems with the same quantum numbers as the dominating boson.
For example, the $\alpha$-parameter for the axial $(\pi\rho)_1$-pair could fixed
at the quark-model prediction 0.40, see Table~\ref{tab.gpair}.
The $BB$-Pair couplings are calculated, assuming unbroken $SU(3)$-symmetry, 
from the $NN$-Pair coupling and the $F/(F+D)$-ratio using $SU(3)$.
So, in addition to the 14 parameters used in Ref.~\cite{RS96ab} we now have
6 pair-coupling fit parameters. 
In Table~\ref{tab.gpair} the fitted pair-couplings are given.
The $(\pi\rho)_1$-coupling is large as expected from $A_1$-saturation, see 
Ref.~\cite{RS96ab}. 
In Table~\ref{tab.gpair} we show the MPE-coupling constants.        
The used  $\alpha =: F/(F+D)$-ratio's for the MPE-couplings are:
$(\pi\eta)$ pairs $\alpha(\{8_s\})=1.0$, 
$(\pi\pi)_1$ pairs $\alpha_V^e(\{8\}_a)=1.0, \alpha_V^m(\{8\}_a)=0.400$, 
and the $(\pi\rho)_1$ pairs $\alpha_A(\{8\}_a)=0.400$. 
The $(\pi\omega)$ pairs $\alpha(\{8_s\})$ has been set equal to
$\alpha_{pv}=0.365$.\\
\noindent {\it Assuming heavy-meson dominance of the meson-pair couplings, 
similarly to the QQM-couplings
all QQ meson-pair couplings get a factor 1/3, i.e. $g_{QQm_1m_2} = G_{BBm_1m_2}/3$.}\\
\noindent In Tables \ref{tab.nnphas3}, \ref{tab.qnphas3} and \ref{tab.qqphas3}
the NN, QN, and QQ phases are shown respectively. This for the 
complete ESC16 model with NN parameters {\bf parbbsc.new17def2}.
The scale parameters for the QQM-couplings and quark mass are
both 1/3. The gluon mass $m_{glu}=420$ MeV/c$^2$, the QQM form factor cut-off is set to 
$\Lambda_{QQM}=986.6$ MeV/c$^2$, and furthermore the confinement 
potential is left out.
Note that the extra couplings for the QQM-vertex w.r.t. the NNM-couplings are set to zero.

\onecolumngrid  ! why not possible?
\begin{table}[hbt]
\caption{ Nucleon-Nucleon ESC16 nuclear-bar $PP$ and $NP$ phases in degrees.}
\begin{ruledtabular}
\begin{tabular}{crrrrrrrrrr} & & & & & &&&&&\\
 $T_{\rm lab}$ & 0.38& 1 & 5  & 10 & 25 & 50 & 100 & 150 & 215 & 320 \\ \hline
     &    &     &     &     &    &&&&& \\
 $^{1}S_{0}(np)$ & 54.57  & 62.02 & 63.47 & 59.72& 50.48    
                 & 39.82  & 25.45 & 15.11 & 4.65 & --8.34  \\
 $^{1}S_{0}$ & 14.62  & 32.62 & 54.75 & 55.16& 48.67    
             & 38.97  & 25.06 & 14.85 & 4.44 &--8.53  \\
 $^{3}S_{1}$ & 159.39 & 147.77& 118.25& 102.72& 80.81  
             & 63.03  & 43.62 & 31.27 & 19.58 & 5.83 \\
 $\epsilon_{1}$ & 0.03  & 0.11 & 0.68 & 1.17 & 1.82   
                & 2.15  & 2.50 & 2.94 & 3.64 & 4.93 \\
 $^{3}P_{0}$ & 0.02   &  0.14 & 1.61 & 3.81 & 8.81    
             & 11.80  &  9.68 & 4.83 &--1.86 &--11.73 \\
 $^{3}P_{1}$ & --0.01  &--0.08  &--0.89  & --2.04  & --4.89     
             & --8.29  &--13.28 &--17.35 & --21.87 & --27.90 \\
 $^{1}P_{1}$ & --0.05  &--0.19  &--1.50  & --3.07  & --6.39     
             & --9.81  &--14.65 &--18.75 & --23.38 & --29.44 \\
 $^{3}P_{2}$ &  0.00  & 0.02  & 0.22  &  0.67  &  2.51     
             &  5.80  & 10.90 & 14.04  &  16.24 &  17.07 \\
 $\epsilon_{2}$ &--0.00  &--0.00 &--0.05 &--0.20 &--0.81    
                &--1.71  &--2.71 &--2.99 &--2.84 &--2.18 \\
 $^{3}D_{1}$ & --0.00  &--0.01  &--0.18  & --0.68  & --2.83    
             &--6.51  &--12.40 &--16.69 & --20.72 & --25.04 \\
 $^{3}D_{2}$ & 0.00  & 0.01  & 0.22  &  0.85  &  3.70     
             & 8.93  & 17.22 & 22.15 &  24.99 &  25.05 \\
 $^{1}D_{2}$ & 0.00  & 0.00  & 0.04  &  0.17  &  0.69     
             & 1.70  & 3.78  & 5.70  &  7.64  &   9.20 \\
 $^{3}D_{3}$ & 0.00  & 0.00  & 0.00  &  0.00  &  0.03    
             & 0.24  & 1.17  & 2.31  &  3.61  &  4.86  \\
 $\epsilon_{3}$ & 0.00  & 0.00 & 0.01 & 0.08 & 0.55   
                & 1.59  & 3.46 & 4.81 & 5.97 & 6.99 \\
 $^{3}F_{2}$ & 0.00  & 0.00  & 0.00  &  0.01  &  0.11     
             & 0.34  & 0.80  & 1.10  &  1.14  &  0.39  \\
 $^{3}F_{3}$ & --0.00  & --0.00  &--0.01  & --0.03  & --0.23     
             &--0.67  &--1.46  &--2.06  & --2.66  & --3.50  \\
 $^{1}F_{3}$ & --0.00  & --0.00  &--0.01  & --0.06  & --0.41     
             &--1.10  &--2.11  &--2.77  & --3.46  & --4.69  \\
 $^{3}F_{4}$ & 0.00  & 0.00  & 0.00  &  0.00  &  0.02     
             & 0.12  & 0.51  & 1.04  &  1.80  &  3.00  \\
 $\epsilon_{4}$ & --0.00  & --0.00 & --0.00 &--0.00 &--0.05    
                &--0.19  &--0.53 &--0.83 &--1.13 &--1.46 \\
 $^{3}G_{3}$ &--0.00 &--0.00  &--0.00  &--0.00  & --0.05    
             &--0.26 &--0.93  &--1.73  &--2.77  & --4.17  \\
 $^{3}G_{4}$ & 0.00 & 0.00  & 0.00  & 0.01  &  0.17     
             & 0.71  & 2.11  & 3.52  &  5.17  &  7.28  \\
 $^{1}G_{4}$ & 0.00 & 0.00  & 0.00  & 0.00  &  0.04     
             & 0.15  & 0.41  & 0.69  &  1.06  &  1.70  \\
 $^{3}G_{5}$ &--0.00 &--0.00  &--0.00  &--0.00  & --0.01      
             &--0.05  &--0.16  &--0.25  & --0.28  & --0.19  \\
 $\epsilon_{5}$ & 0.00 & 0.00  & 0.00 & 0.00 & 0.04    
                & 0.20  & 0.70 & 1.22 & 1.83 & 2.62 \\
     &    &     &     &     &  &&&&&  \\
\end{tabular}
\end{ruledtabular}
\label{tab.nnphas3}   
\end{table}
\begin{table}[hbt]
\caption{ Quark-Nucleon ESC16 nuclear-bar $UP$ and $DP$ phases in degrees.}
\begin{ruledtabular}
\begin{tabular}{crrrrrrrrrr} & & & & & &&&&&\\
 $T_{\rm lab}$ & 0.38& 1 & 5  & 10 & 25 & 50 & 100 & 150 & 215 & 320 \\ \hline
     &    &     &     &     &    &&&&& \\
	$^{1}S_{0}(np)$ &  0.288 &  0.445&  0.727&  0.648& --0.254 
		 &--2.278&--6.443&--10.293&--14.761&--20.894\\
$^{1}S_{0}$ &  0.194 &  0.364&  0.700&  0.650 &--2135    
	     &--2.207 &--6.348&--10.192 &--14.661&--20.804\\
	$^{3}S_{1}$ &   0.544&   0.855 &  1.614 & 1.865 & 1.559 
             &   0.828&--2.159&--6.54 &--10.52&--16.09 \\
 $\epsilon_{1}$ & 0.007 & 0.030& 0.260& 0.568& 1.254  
                & 1.828 & 2.159& 2.153& 2.021& 1.765\\
 $^{3}P_{0}$ & 0.003  &  0.015& 0.163& 0.379& 0.885   
	     &  1.230 &  0.926& 0.052&--1.435&--4.205 \\
 $^{3}P_{1}$ &--0.002  &--0.011 &--0.120 &--0.283  &--0.707     
             &--1.180  &--1.677 &--1.922 &--2.159  &--2.699  \\
 $^{1}P_{1}$ & --0.000 &--0.000 &--0.000 & --0.393 & --0.830    
             & --1.141 &--1.174 &--0.964 & --0.680 & --0.568 \\
 $^{3}P_{2}$ &  0.000 & 0.002 & 0.029 &  0.085 &  0.333    
             &  0.857 &  1.977& 2.990  &  3.985 &   4.720\\
 $\epsilon_{2}$ &--0.000 &--0.000&--0.005&--0.020&--0.089   
                &--0.203 &--0.346&--0.402&--0.395&--0.282\\
 $^{3}D_{1}$ & --0.000 &--0.000 &--0.010 & --0.042 & --0.210   
             &--0.515 &--0.864 &--0.871 & --0.567 &   0.190 \\
 $^{3}D_{2}$ & 0.000 & 0.000 & 0.000 &  0.000 &  0.000    
             &  0.893&  1.755&  2.343&   2.806&   3.068\\
 $^{1}D_{2}$ & 0.000 & 0.000 & 0.000 &  0.000 &  0.000    
             & 0.201 & 0.468 & 0.741 &  1.068 &   1.443\\
 $^{3}D_{3}$ & 0.000 & 0.000 & 0.000 &  0.000 &  0.000   
             & 0.000 & 0.000 &  0.233&  0.538 &  1.161 \\
 $\epsilon_{3}$ & 0.000 & 0.000& 0.000& 0.000 & 0.000  
                & 0.000 & 0.000& 0.583& 0.721& 0.796\\
 $^{3}F_{2}$ & 0.000 & 0.000 & 0.000 &  0.000 &  0.000    
             & 0.033 & 0.076 & 0.086 &  0.042 &--0.121 \\
 $^{3}F_{3}$ &   0.000 &   0.000 &  0.000 &   0.000 &   0.000    
             &- 0.000 &  0.000 &--0.281 & --0.371 & --0.442 \\
 $^{1}F_{3}$ & --0.000 & --0.000 &--0.000 & --0.000 & --0.000    
             &--0.000 &--0.000 &--0.410 & --0.543 & --0.728 \\
 $^{3}F_{4}$ & 0.000 & 0.000 & 0.000 &  0.000 &  0.000    
             & 0.000 & 0.000 & 0.000 &  0.127 &  0.233 \\
 $\epsilon_{4}$ & --0.000 & --0.000& --0.000&--0.000&--0.000   
		&--0.000 &--0.000&--0.000&--0.129&--0.165\\
 $^{3}G_{3}$ &--0.000&--0.000 &--0.000 &--0.000 & --0.000   
             &--0.000&--0.000 &--0.142 &--0.199 & --0.213 \\
 $^{3}G_{4}$ &   0.000&   0.000 &   0.000 &   0.000 &    0.000    
             &   0.000 &   0.000 &   0.000 &    0.000 &   0.728\\
 $^{1}G_{4}$ & 0.000& 0.000 & 0.000 & 0.000 &  0.000    
             & 0.000 & 0.000 & 0.000 &  0.112 &  0.179 \\
 $^{3}G_{5}$ &--0.000&--0.000 &--0.000 &--0.000 & --0.000     
             &--0.000 &--0.000 &--0.000 & --0.000 & --0.000 \\
 $\epsilon_{5}$ & 0.000& 0.000 & 0.000& 0.000& 0.000   
                & 0.000 & 0.000& 0.000& 0.000& 0.0000\\
     &    &     &     &     &  &&&&&  \\
\end{tabular}
\end{ruledtabular}
\label{tab.qnphas3}   
\end{table}
\begin{table}[hbt]
\caption{ Quark-Quark ESC16 nuclear-bar $UU$ and $DU$ phases in degrees.}
\begin{ruledtabular}
\begin{tabular}{crrrrrrrrrr} & & & & & &&&&&\\
 $T_{\rm lab}$ & 0.38& 1 & 5  & 10 & 25 & 50 & 100 & 150 & 215 & 320 \\ \hline
     &    &     &     &     &    &&&&& \\
 $^{1}S_{0}(np)$ &  0.864 & 1.391 & 3.047 & 4.203& 6.197    
		 &--7.869 &--9.103&--9.182&--8.529& --6.686 \\
 $^{1}S_{0}$ & 1.099  & 2.168 & 5.665 & 8.144& 12.452   
             & 16.147 & 19.268& 20.221&20.155& 18.781 \\
 $^{3}S_{1}$ &  0.688 &  1.109&  2.440&  3.384& 5.071  
             & --6.604&--7.982&--8.360&--8.115&-6.853 \\
 $\epsilon_{1}$ & 0.001 & 0.002& 0.024& 0.060& 0.175  
                & 0.338 & 0.569& 0.733& 0.900& 1.123 \\
 $^{3}P_{0}$ & 0.001  &  0.004& 0.050& 0.142& 5.25    
             &  1.330 & 3.106 &4.805 & 6.688& 8.847  \\
 $^{3}P_{1}$ &  0.000  &  0.001 & 0.010  &  0.033  &  0.152     
             &  0.473  & 1.382  &  2.466 &  3.964  &  6.328  \\
 $^{1}P_{1}$ & --0.000 &--0.000 &--0.000 & --0.035 & --0.093    
             & --0.156 &--0.186 &--0.145 & --0.038 &  0.176  \\
 $^{3}P_{2}$ &  0.000 & 0.001 & 0.017 &  0.049 &  0.197    
             &  0.559 &  1.551& 2.779  &  4.600 &  7.813 \\
 $\epsilon_{2}$ &--0.000 &--0.000&--0.000&--0.001&--0.006   
                &--0.021 &--0.057&--0.096&--0.147&--0.213\\
 $^{3}D_{1}$ & --0.000 &--0.000 &--0.000 & --0.002 & --0.011   
             &--0.036 &--0.079 &--0.088 & --0.039 &   0.185 \\
 $^{3}D_{2}$ & 0.000 & 0.000 & 0.000 &  0.000 &  0.000    
             &  0.060&  0.145&  0.219&   0.303&   0.428\\
 $^{1}D_{2}$ & 0.000 & 0.000 & 0.000 &  0.000 &  0.000    
             & 0.022 & 0.080 & 0.173 &  0.345 &   0.743\\
 $^{3}D_{3}$ & 0.000 & 0.000 & 0.000 &  0.000 &  0.000   
             & 0.000 & 0.000 &--0.026&--0.019 &  0.066 \\
 $\epsilon_{3}$ & 0.000 & 0.000& 0.000& 0.000 & 0.000  
                & 0.000 & 0.000& 0.050& 0.073& 0.099\\
 $^{3}F_{2}$ & 0.000 & 0.000 & 0.000 &  0.000 &  0.000    
             & 0.003 & 0.016 & 0.043 &  0.106 &  0.280 \\
 $^{3}F_{3}$ &   0.000 &   0.000 &  0.000 &   0.000 &   0.000    
             &- 0.000 &  0.000 &--0.008 &   0.006 &   0.068 \\
 $^{1}F_{3}$ & --0.000 & --0.000 &--0.000 & --0.000 & --0.000    
             &--0.000 &--0.000 &--0.037 & --0.052 & --0.066 \\
 $^{3}F_{4}$ & 0.000 & 0.000 & 0.000 &  0.000 &  0.000    
             & 0.000 & 0.000 & 0.000 &  0.012 &  0.027 \\
 $\epsilon_{4}$ & --0.000 & --0.000& --0.000&--0.000&--0.000   
		&--0.000 &--0.000&--0.000&--0.011&--0.018\\
 $^{3}G_{3}$ &--0.000&--0.000 &--0.000 &--0.000 & --0.000   
             &--0.000&--0.000 &--0.008 &--0.016 & --0.031 \\
 $^{3}G_{4}$ &   0.000&   0.000 &   0.000 &   0.000 &    0.000    
             &   0.000 &   0.000 &   0.000 &    0.000 &   0.053\\
 $^{1}G_{4}$ & 0.000& 0.000 & 0.000 & 0.000 &  0.000    
             & 0.000 & 0.000 & 0.000 &  0.010 &  0.019 \\
 $^{3}G_{5}$ &  0.000&  0.000 &  0.000 &  0.000 &   0.000     
             &  0.000 &  0.000 &  0.000 &   0.000 &   0.000 \\
 $\epsilon_{5}$ & 0.000& 0.000 & 0.000& 0.000& 0.000   
                & 0.00R0& 0.000& 0.000& 0.000& 0.000\\
     &    &     &     &     &  &&&&&  \\
\end{tabular}
\end{ruledtabular}
\label{tab.qqphas3}   
\end{table}
\onecolumngrid  
 \begin{table}[hbt]
\caption{ESC16 Low energy parameters: S-wave scattering lengths and 
effective ranges, deuteron binding energy $E_B$, and electric 
quadrupole $Q_e$.
Experimental values and references, see \cite{NPB147}.
The asterisk denotes that the low-energy parameters were not searched.}  
\begin{center}
\begin{ruledtabular}    
\begin{tabular}{ccccc} & & & & \\
     & \multicolumn{3}{c}{experimental data}& ESC16
     \\ &&&& \\ \hline
 $a_{pp}(^1S_0)$ & --7.828 & $\pm$ & 0.008 & --7.7718\\
 $r_{pp}(^1S_0)$ &  \hspace{1mm} 2.800 & $\pm$ & 0.020 &\hspace{2mm} 2.7612$^\ast$ \\ \hline
 $a_{np}(^1S_0)$ & --23.748 & $\pm$ & 0.010 & --23.7346\\
 $r_{np}(^1S_0)$ &\hspace{2mm}   2.750 & $\pm$ & 0.050 &\hspace{2mm} 2.6992$^\ast$ \\ \hline
 $a_{nn}(^1S_0)$ & --18.63     & $\pm$ & 0.48  & --17.783\\
 $r_{nn}(^1S_0)$ &\hspace{1mm} 2.860   & $\pm$ & 0.15  &\hspace{2mm} 2.8301$^\ast$ \\ \hline
 $a_{np}(^3S_1)$ &\hspace{1mm} 5.424 & $\pm$ & 0.004 &\hspace{2mm} 5.4396$^\ast$ \\
 $r_{np}(^3S_1)$ &\hspace{1mm} 1.760 & $\pm$ & 0.005 &\hspace{2mm} 1.7488$^\ast$\\ \hline
  $E_B$         &  --2.224644 & $\pm$ & 0.000046 & --2.224636 \\
  $Q_e$         &\hspace{1mm} 0.286 & $\pm$ & 0.002 &\hspace{1mm} 0.2727 \\ 
\end{tabular}
\end{ruledtabular}   
\end{center}
 \label{tab.lowenergy}
 \end{table}

\section{Summary and Outlook}      
\label{sec:discussion}
The ESC-approach to the baryon-baryon interactions is able to make a connection between 
the  available baryon-baryon data on the one hand, and on the other hand the underlying
quark structure of the baryons and mesons. Namely, a succesfull description
of both the $N\!N$- and $Y\!N$-scattering data is obtained with meson-baryon
coupling parameters which are almost all explained by the QPC-model, 
which implicitly makes use of the CQM.
The finding that in the CQM it is possible to derive the ESC baryon-baryon
meson-exchange potentials from meson-exchange between quarks via folding 
with the ground-state baryon quark wave functions opens the way to derive
meson-exchange quark-quark potentials almost parameter free.

The method followed in this paper is based on these observations. 
The potentials are worked out in a $1/m_Q$-expansion. 
For quark masses significantly smaller than the constituent quarks the Kadyshevsky formalism in
momentum space provides a suitable framework for relativistic calculations. In this case    
the $1/m_Q$-expansion can be avoided by using the complete formulas for the Kadyshevsky diagrams.
This lowering of the quark mass will happen in dense quark-matter, and therefore
a relativistic many-body theory is eventually needed. Similar to the Dirac-Bruckner Theory,
the Kadyshevsky-Bethe-Goldstone equation for the G-matrix is obtained in momentum-space, 
which can be solved using standard methods.

\noindent {\it The tables for the Quark-quark and Quark-nucleon phase shifts
have some physical reality only for the hypothetical case of completely deconfined matter.
They mainly serve to give an impression of the strength of the quark-quark and 
quark-nucleon interactions compared to the nucleon-nucleon interactions.}

\noindent Application of this work is the recent study of neutron-star (NS) matter
modeled as a mixture of quark and baryon matter. The G-matrices of both kinds
of matter are described with largely common parameters \cite{YYR22,YYR23,YYR24}.\\
\noindent Finally, we mention the possibility to derive an $\Omega\Omega$-potential by 
folding the QQ-potentials with the $\Omega$ three-quark wave function.


 \appendix


\onecolumngrid
 \begin{figure}   
\begin{flushleft}
\rule{16cm}{0.5mm}
\end{flushleft}
	 \vspace{-2.5cm}
  \resizebox{!}{!}
 {\includegraphics[210,650][410,900]{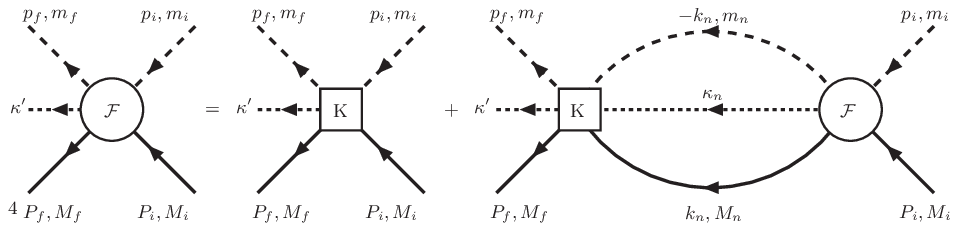}}
  \caption{\sl G-matrix: Kadyshevsky-Bethe-Goldstone Equation}                         
  \label{fig:inteq6} 
 \end{figure}
\section{ Kadyshevsky G-matrix Equation}    
\label{app:kadgmat}         
In a fermi-system, {\it e.g.} quark matter, 
the Kadyshevsky-Bethe-Goldstone-Kadyshevsky equation (BGKE) is
depicted in Fig.~\ref{fig:inteq6}, and reads
\begin{eqnarray}
&& {\cal F}({\bf p'},{\bf p};W) = K({\bf p'},{\bf p};W) +         
 \frac{1}{(2\pi)^3}\int \frac{d^3p^{\prime\prime}}
 {2E({\bf p}^{\prime\prime})}\ K({\bf p'},{\bf p}^{\prime\prime};W)\ 
\cdot\nonumber\\ && \times
\frac{M^2}{E({\bf p}^{\prime\prime})\left[E({\bf p}^{\prime\prime})-
 E({\bf p})-i\epsilon\right]}\
  Q_P[n_F(p'')]\ {\cal F}({\bf p}^{\prime\prime},{\bf p};W), \nonumber\\
\label{app:gmat.a}\end{eqnarray}
which corresponds to Eq.~(\ref{eq:4.11}). 
Then, the Bethe-Goldstone-Kadyshevsky two-particle wave function reads
\begin{eqnarray}
&& \psi(p;W) = \psi^{(0)}(p) + \int \frac{d^3p^{\prime\prime}}
 {2E({\bf p}^{\prime\prime})(2\pi)^3}\cdot\nonumber\\ && \times
\frac{M^2}{E({\bf p}^{\prime\prime})\left[E({\bf p}^{\prime\prime})-
 E({\bf p})-i\epsilon\right]}\
  Q_P[n_F(p'')]\ \psi(p'';W),            
\label{app:gmat.b}\end{eqnarray}
where $\psi^{(0)}(p;W)$ corresponds to the two-particle plane-wave product state
$|\phi_0(p_1)\rangle|\phi_0(p_2)\rangle$, with 
$P=p_1+p_2, p=p_1-p_2$, and $W=p_1^0+p_2^0$. Here, $\phi^{(0)}(p)$ is the
plane wave in the case of matter or a model wave function for finite nuclei.\\
Then, the corresponding G-matrix is introduced in the standard way by defining           
 $ {\cal G}(p;W) = \langle \psi^{(0)}| K_{op}\ |\psi(p;W)\rangle$,  giving the equation
\begin{eqnarray}
&& {\cal G}({\bf p'},{\bf p};W) = K({\bf p'},{\bf p};W) +         
 \frac{1}{(2\pi)^3}\int \frac{d^3p^{\prime\prime}}
 {2E({\bf p}^{\prime\prime})}\
 K({\bf p'},{\bf p}^{\prime\prime};W) 
\cdot\nonumber\\ && \times 
\frac{M^2}{E({\bf p}^{\prime\prime})\left[E({\bf p}^{\prime\prime})-
 E({\bf p})\right]}\
  Q_P[n_F(p'')]\ {\cal G}({\bf p}^{\prime\prime},{\bf p};W). \nonumber\\
\label{app:gmat.c}\end{eqnarray}
This integral equation for the G-matrix is similar to that in the 
Dirac-Bruckner theory, see {\it e.g.} \cite{BTH87,Brock90}. 
Notice that in the non-relativistic limit $M/E=1$ and Eqn.~(\ref{app:gmat.c}) 
corresponds to the usual employed G-matrix equation in the many-body problem.
In fact, the difference with Eqn.~(1) of Refs.~\cite{Baldo91,Burgio02} is largely 
a factor $(M/E(p''))^2$ under the integral, and the use of an effective density
dependent mass in the Dirac spinors.  
Therefore, the momentum space evaluation of the G-matrix partial waves is wel known.

\noindent For a quark pair with flavor quantum numbers $f_1,f_2$ in quark matter        
the G-matrix equation for partial waves in short notation reads
\begin{eqnarray}
	G_{cc_0}(\omega) &=& K_{cc_0} + \sum_{c'}\ 
	\left[\frac{m_Q}{(\epsilon_{f_1'}+\epsilon_{f'_2})}\right]^2
	K_{cc'}\
	\frac{Q_{y'}}{\omega-\epsilon_{f'_1}-\epsilon_{f'_2} }\ G_{c'c_0}(\omega),
\label{app:gmat.d}\end{eqnarray}
where c denotes the 'relative' state $(y,T,L,S,J)$ with $y= (f_1,f_2)$. S and T 
are spin and isospin quantum numbers, respectively. The energies are
$\epsilon_{f_i} = \sqrt{k_{f_i}^2+m_Q^2}-m_Q$, i=1,2.
The quark single particle (s.p.) energy $\epsilon_f$ in quark matter is
\begin{equation}
	\epsilon_f(k_f) = \left[\sqrt{k_f^2+m_Q^2}]-m_Q\right] + U_f(k_f), 
\label{app:gmat.e}\end{equation}  
where $k_f$ is the f-quark momentum ($\hbar = c=1$). The potential energy
$U_f$ is (ontained self-consistently) in terms of the G-matrix as
\begin{eqnarray}
 U_f(k_f) &=& \sum_{|{\bf k}_{f'}|} \langle {\bf k}_f{\bf k}_{f'} |
	G_{ff'}\left(\omega = \epsilon_f(k_f)+\epsilon_{f'}(k_{f'}\right)
	|{\bf k}_f{\bf k}_{f'}\rangle.
\label{app:gmat.f}\end{eqnarray}  
The kinetic, potential, and total energies per quark are given by averaged
quantities of $T_f, U_f$, and $E_f= T_f+U_f$ in a Fermi sphere.  

\onecolumngrid
\section{ BBM-couplings in the QPC-model}   
\label{app:qpc}         
The BBM-couplings in the ESC models fit very well with the 
$^3P_0 \oplus ^3S_1$ quark-pair creation (QPC) model.  
A simple (effective) QPC interaction Lagrangian is                      
\begin{equation}  
 {\cal L}_I = \gamma \left[A \left(\sum_{j} \bar{q}_j\ q_j\right)\cdot
                     \left(\sum_{i} \bar{q}_i\ q_i\right)
            + B \left(\sum_{j} \bar{q}_j \gamma_\mu\ q_j\right)\cdot
                 \left(\sum_{i} \bar{q}_i \gamma^\mu\ q_i\right)\right],
\label{app:qpc.1}\end{equation}
where $\gamma$, A, and  B are given in Ref.~\cite{NRY19a} Table II.                                     
\noindent To see the meson couplings we make the Fierz transformation
of (\ref{app:qpc.1}) which gives \cite{Okun84}
\begin{eqnarray}
 {\cal L}_I &=& -\frac{\gamma}{4}\sum_{i,j} \left[ \vphantom{\frac{A}{A}}
 \hspace*{0.0cm}  (A+4B)\bar{q}_i\ q_j\cdot\bar{q}_j\ q_i 
   +(A-4B)\bar{q}_i\gamma_5 q_j\cdot\bar{q}_j\gamma^5 q_i  
\right. \nonumber\\[-1mm] && \left.  \hspace*{1.3cm} 
    +(A-2B)\bar{q}_i\gamma_\mu q_j\cdot\bar{q}_j\gamma^\mu q_i 
    -(A+B)\bar{q}_i\gamma_\mu\gamma_5 q_j\cdot\bar{q}_j\gamma^\mu\gamma^5 q_i 
\right. \nonumber\\[-0mm] && \left.  \hspace*{1.3cm} 
     -(A/2)\ \bar{q}_i\sigma_{\mu\nu} q_j\cdot 
     \bar{q}_j\sigma^{\mu\nu} q_i \vphantom{\frac{A}{A}}\right].    
\label{app:qpc.2}\end{eqnarray}
Identifying the $\bar{q} q$ pairs with the mesons
\begin{eqnarray}
  \chi^S_{ij} &\sim& \bar{q}_j\ q_i\ ,\ 
  \chi^P_{ij} \sim \bar{q}_j\gamma_5\ q_i\ ,\ 
  \chi^V_{\mu,ij} \sim \bar{q}_j\gamma_\mu q_i\ ,\ 
  \chi^A_{\mu,ij} \sim \bar{q}_j\gamma_5\gamma_\mu q_i 
\label{app:qpc.3}\end{eqnarray}
the QQM-couplings are defined. For example, the pseudoscalar couplings
are 
\begin{equation}
 {\cal H}_P = g^{(p)}_8\sqrt{2}\left[\bar{Q} M_P^{(8)} Q\right]
 + g^{(p)}_1 \left[\bar{Q}M_P^{(1)}Q\right]/\sqrt{3}, 
\label{app:qpc.4}\end{equation}
where $g^{(p)}_8 = -\gamma_P (A-4B)/4$.

\onecolumngrid


\section{Momentum-space Meson-Quark-Quark Vertices }
\label{app:MQV}
\subsection{Pauli-reduction Dirac-spinor $\Gamma$-matrix elements}
\label{app:Pauli}
The transition from Dirac spinors to Pauli spinors is given here,  
without approximations. We use the notations ${\cal E}= E+M$ and
${\cal E}'= E'+M'$, where $E=E(p,M)$ and $E'=E(p',M')$.
Also, we omit, on the right-hand side in the expressions below,
the final and initial Pauli spinors $\chi^{\prime \dagger}$ and
$\chi$ respectively, which are self-evident.
\begin{subequations}
\label{Pauli.1}
\begin{eqnarray}
   \bar{u}({\bf p}') u({\bf p}) &=&
 +\sqrt{\frac{{\cal E}'{\cal E}}{4M' M}}
  \left[ \left(1-\frac{{\bf p}'\cdot{\bf p}}{{\cal E}'{\cal E}}\right)
 -i\frac{ {\bf p}'\times{\bf p}\cdot\mbox{\boldmath $\sigma$}}{{\cal E}'{\cal E}} 
\right]
     ,    \label{Paus}\\
   \bar{u}({\bf p}')\gamma_5 u({\bf p}) &=&
 -\sqrt{\frac{{\cal E}'{\cal E}}{4M' M}}
 \left[ \frac{\mbox{\boldmath $\sigma$}\!\cdot\!{\bf p}'}{{\cal E}'}
      - \frac{\mbox{\boldmath $\sigma$}\!\cdot\!{\bf p}}{{\cal E}} \right]
     ,                      \label{Pau1p}\\
   \bar{u}({\bf p}')\gamma^{0} u({\bf p}) &=&
 +\sqrt{\frac{{\cal E}'{\cal E}}{4M' M}}
   \left[ \left(1+\frac{{\bf p}'\cdot{\bf p}}{{\cal E}'{\cal E}}\right)
 +i \frac{{\bf p}'\times{\bf p}\cdot\mbox{\boldmath $\sigma$}}{{\cal E}'{\cal E}}\right]
     ,    \label{Pauv0}\\
   \bar{u}({\bf p}')\mbox{\boldmath $\gamma$}\ u({\bf p}) &=& 
 +\sqrt{\frac{{\cal E}'{\cal E}}{4M' M}}
\left[ \left(\frac{{\bf p}'}{{\cal E}'}+\frac{{\bf p}}{{\cal E}}\right) 
 + i \left(\frac{\mbox{\boldmath $\sigma$}\times{\bf p}'}{{\cal E}'}
         - \frac{\mbox{\boldmath $\sigma$}\times{\bf p}}{{\cal E}}\right)
 \right] , \label{Pauvi}\\
   \bar{u}({\bf p}')\gamma_5\gamma^{0} u({\bf p}) &=&
 -\sqrt{\frac{{\cal E}'{\cal E}}{4M' M}}
 \left[ \frac{\mbox{\boldmath $\sigma$}\!\cdot\!({\bf p}'}{{\cal E}'}
      + \frac{\mbox{\boldmath $\sigma$}\!\cdot\!({\bf p}}{{\cal E}} \right]
     ,    \label{Paua0}\\
   \bar{u}({\bf p}')\gamma_5\mbox{\boldmath $\gamma$}\ u({\bf p}) &=& 
 -\sqrt{\frac{{\cal E}'{\cal E}}{4M' M}}
  \left[ \mbox{\boldmath $\sigma$} + \frac{ 
  (\mbox{\boldmath $\sigma$}\cdot{\bf p}')\ \mbox{\boldmath $\sigma$}\ 
  (\mbox{\boldmath $\sigma$}\cdot{\bf p})}{{\cal E}'{\cal E}}\right]  
 \nonumber\\ &=& 
 -\sqrt{\frac{{\cal E}'{\cal E}}{4M' M}}\left[
  \left(1-\frac{{\bf p}'\cdot{\bf p}}{{\cal E}'{\cal E}}\right)\mbox{\boldmath $\sigma$} 
 -i \frac{{\bf p}'\times{\bf p}}{{\cal E}'{\cal E}} 
 \right.\nonumber\\ && \left.
 +\frac{1}{{\cal E}'{\cal E}}\left(
 \mbox{\boldmath $\sigma$}\cdot{\bf p}\ {\bf p}' + 
  \mbox{\boldmath $\sigma$}\cdot{\bf p}'\ {\bf p}\right)\right] \approx 
  -\mbox{\boldmath $\sigma$} 
 , \label{Pauai} ,
\end{eqnarray}
\end{subequations}
\noindent
$\!\!\!$where we defined ${\bf k}={\bf p}'-{\bf p}$,
${\bf q}=({\bf p}'+{\bf p})/2$, and $\kappa_{V}=f_{V}/g_{V}$.\\

Using the the Gordon decomposition
\begin{eqnarray}
&& i\ \bar{u}(p')\ \sigma^{\mu\nu}(p'-p)_\nu u(p) = \bar{u}(p')\left\{\vphantom{\frac{A}{A}}
 (M'+M)\gamma^\mu - (p'+p)^\mu\right\}\ u(p)
\label{Gordon.app}\end{eqnarray}
one obtains for the complete vector-vertex 
\begin{subequations}
\label{Paul.3}
\begin{eqnarray}
 \bar{u}(p') \Gamma^\mu_V u(p) &\equiv& \bar{u}(p')\left[\gamma^\mu 
 + \frac{i}{2{\cal M}}\kappa_V \sigma^{\mu\nu}(p'-p)_\nu\right] u(p)                  
 \nonumber\\[3mm]           
 &=& \bar{u}(p')\left[\left(1+\frac{M'+M}{2{\cal M}}\kappa_V\right)\gamma^\mu 
 - \frac{\kappa_V}{2{\cal M}}(p'+p)_\mu\right] u(p) \Longrightarrow 
 \nonumber\\[3mm]           
 \mu=0 &:& +\sqrt{\frac{{\cal E}'{\cal E}}{4M' M}}\left[
  \left(1+\frac{M'+M}{2{\cal M}}\kappa_V\right)
 \left(1+\frac{\mbox{\boldmath $\sigma$}\cdot{\bf p}'\
 \mbox{\boldmath $\sigma$}\cdot{\bf p}}{{\cal E}'{\cal E}}\right)
 \right.\nonumber\\ && \left. 
 -\frac{\kappa_V}{2{\cal M}}(E'+E) 
 \left(1-\frac{\mbox{\boldmath $\sigma$}\cdot{\bf p}'\
 \mbox{\boldmath $\sigma$}\cdot{\bf p}}{{\cal E}'{\cal E}}\right)\right]  
, \label{PAU0}\\
  \mu=i &:& +\sqrt{\frac{{\cal E}'{\cal E}}{4M' M}}\left[
 \left(1+\frac{M'+M}{2{\cal M}}\kappa_V\right)
 \left\{\left(\frac{{\bf p}'}{{\cal E}'}+\frac{\bf p}{\cal E}\right)
 +i\left(\frac{\mbox{\boldmath $\sigma$}\times{\bf p}'}{{\cal E}'}
 - \frac{\mbox{\boldmath $\sigma$}\times{\bf p}}{{\cal E}}\right)\right\}
 \right.\nonumber\\ && \left.
 -\frac{\kappa_V}{2{\cal M}}({\bf p}'+{\bf p}) 
 \left(1-\frac{\mbox{\boldmath $\sigma$}\cdot{\bf p}'\
 \mbox{\boldmath $\sigma$}\cdot{\bf p}}{{\cal E}'{\cal E}}\right)\right]  
. \label{PAUi} 
\end{eqnarray}
\end{subequations}
 
\begin{flushleft}
\rule{16cm}{0.5mm}
\end{flushleft}
\subsection{1/M-expansion $\Gamma$-matrix elements}

The exact transition from Dirac spinors to Pauli spinors is given in
Appendix~\ref{app:Pauli}. From the expressions in \ref{app:Pauli}, keeping only
terms up to order $1/M$, and setting the scaling mass ${\cal M}=M$, 
we find that the vertex operators in
Pauli-spinor space for the $N\!Nm$ vertices are given by
\begin{subequations}
\label{Pauliexp.3}
\begin{eqnarray}
   \bar{u}({\bf p}') u({\bf p}) &=&
      \left[ \left(1-\frac{{\bf p}'\cdot{\bf p}}{4M^2}\right)
 -\frac{i}{4M^2} {\bf p}'\times{\bf p}\cdot\mbox{\boldmath $\sigma$}\right]
     ,    \label{Gams}\\
   \bar{u}({\bf p}')\gamma_5 u({\bf p}) &=&
 -\frac{1}{2M}\!\left[ \mbox{\boldmath $\sigma$}\!\cdot\!({\bf p}'-{\bf p})\right]
  = -\frac{1}{2M}\!\left[ \mbox{\boldmath $\sigma$}\!\cdot\!{\bf k}\right]
     ,                      \label{Gam1pp}\\
   \bar{u}({\bf p}')\gamma^{0} u({\bf p}) &=&
      \left[ \left(1+\frac{{\bf p}'\cdot{\bf p}}{4M^2}\right)
 +\frac{i}{4M^2} {\bf p}'\times{\bf p}\cdot\mbox{\boldmath $\sigma$}\right]
     ,    \label{Gamv0}\\
   \bar{u}({\bf p}')\mbox{\boldmath $\gamma$}\ u({\bf p}) &=& 
 \frac{1}{2M}\left[ ({\bf p}'+{\bf p}) + i \mbox{\boldmath $\sigma$}\times({\bf p}'-{\bf p})
 \right] , \label{Gamvi}\\
   \bar{u}({\bf p}')\gamma_5\gamma^{0} u({\bf p}) &=&
 -\frac{1}{2M}\!\left[ \mbox{\boldmath $\sigma$}\!\cdot\!({\bf p}'+{\bf p})\right]
 = -\frac{1}{M}\!\left[ \mbox{\boldmath $\sigma$}\!\cdot\!{\bf q}\right]
     ,    \label{Gama0}\\
   \bar{u}({\bf p}')\gamma_5\mbox{\boldmath $\gamma$}\ u({\bf p}) &=& 
 -\left[ \mbox{\boldmath $\sigma$} + \frac{1}{4M^2} 
  (\mbox{\boldmath $\sigma$}\cdot{\bf p}')\ \mbox{\boldmath $\sigma$}\ 
  (\mbox{\boldmath $\sigma$}\cdot{\bf p})\right]  
 = -\left[\left(1-\frac{{\bf p}'\cdot{\bf p}}{4M^2}\right)\mbox{\boldmath $\sigma$}
 \right.\nonumber\\ && \left.
 -\frac{i}{4M^2} {\bf p}'\times{\bf p} +\frac{1}{4M^2}\left(
 \mbox{\boldmath $\sigma$}\cdot{\bf p}\ {\bf p}' + 
  \mbox{\boldmath $\sigma$}\cdot{\bf p}'\ {\bf p}\right)\right] \approx 
  -\mbox{\boldmath $\sigma$} 
 , \label{Gamai} ,
\end{eqnarray}
\end{subequations}
\noindent
$\!\!\!$where we defined ${\bf k}={\bf p}'-{\bf p}$,
${\bf q}=({\bf p}'+{\bf p})/2$, and $\kappa_{V}=f_{V}/g_{V}$.
In passing we note that the inclusion of the $1/M^2$-terms is necessary
in order to get spin-orbit potentials, like in the case of the OBE-potentials.\\

For the magnetic-coupling we use the Gordon decomposition
\begin{eqnarray}
&& i\ \bar{u}(p')\ \sigma^{\mu\nu}(p'-p)_\nu u(p) = \bar{u}(p')\left\{\vphantom{\frac{A}{A}}
 2M \gamma^\mu - (p'+p)^\mu\right\}\ u(p)
\label{Gordon}\end{eqnarray}
We get 
\begin{subequations}
\label{Pauliexp.4}
\begin{eqnarray}
&& i\ \bar{u}(p')\ \sigma^{\mu\nu}(p'-p)_\nu u(p) \Longrightarrow 
 \nonumber\\[3mm]           
&& \mu=0\ :\ -M \left[ \left(1-\frac{{\bf p}'\cdot{\bf p}}{4M^2}\right)
      +\frac{(p^{\prime 2}+p^2)}{2M^2}
 -\frac{i}{4M^2} {\bf p}'\times{\bf p}\cdot\mbox{\boldmath $\sigma$}\right]
, \label{Gamt0}\\
&& \mu=i\ :\ -\left[ \frac{1}{2}({\bf p}'+{\bf p}) -\frac{i}{2}\mbox{\boldmath $\sigma$}
 \times({\bf p}'-{\bf p})\right] . \label{Gamti}
\end{eqnarray}
\end{subequations}
For the vector-vertex with direct and derivative coupling one has  
\begin{subequations}
\label{Pauliexp.5}
\begin{eqnarray}
 \bar{u}(p') \Gamma^\mu_V u(p) &\equiv& \bar{u}(p')\left[\gamma^\mu 
 + \frac{i}{2M}\kappa_V \sigma^{\mu\nu}(p'-p)_\nu\right] u(p)                  
 \nonumber\\[3mm]           
 &=& \bar{u}(p')\left[(1+\kappa_V)\gamma^\mu 
 - \frac{\kappa_V}{2M}(p'+p)_\mu\right] u(p) \Longrightarrow 
 \nonumber\\[3mm]           
 \mu=0 &:& \left[ (1+\kappa_V)\left(1+\frac{{\bf p}'\cdot{\bf p}}{4M^2}
 +\frac{i}{4M^2}{\bf p}'\times{\bf p}\cdot\bm{\sigma}\right)
\right.\nonumber\\ && \left. 
    -\kappa_V\frac{E_{p'}+E_p}{2M}\left(1-\frac{{\bf p}'\cdot{\bf p}}{4M^2}
 -\frac{i}{4M^2}{\bf p}'\times{\bf p}\cdot\bm{\sigma}\right)\right] \approx
 \nonumber\\ && 
 \left[ 1+(1+2\kappa_V)\left\{\frac{{\bf p}'\cdot{\bf p}}{4M^2}
 +\frac{i}{4M^2}{\bf p}'\times{\bf p}\cdot\bm{\sigma}\right\} 
    -\kappa_V\frac{{\bf p}^{\prime 2}+{\bf p}^2}{4M^2}\right] 
, \label{GAM0}\\
  \mu=i &:& \frac{1}{M}\left[ \frac{1}{2}({\bf p}'+{\bf p}) 
  +\frac{i}{2}(1+\kappa_V)\mbox{\boldmath $\sigma$}\times({\bf p}'-{\bf p})\right] 
. \label{GAMi} 
\end{eqnarray}
\end{subequations}

\subsection{Complete Meson-vertices in Pauli-spinor space}
The transition from Dirac spinors to Pauli spinors is reviewed in
Appendix C of \cite{Rij91}. Following this reference and keeping only
terms up to order $(1/M)^2$, we find that the vertex operators in
Pauli-spinor space for the $Q\!Qm$ vertices are given by

 
\begin{subequations}
\label{Pauliexp.1}
\begin{eqnarray}
   \bar{u}({\bf p}')\Gamma^{(1)}_{P}u({\bf p}) &=&
 -i\frac{f_{P}}{m_{\pi}}\!\left[ \mbox{\boldmath $\sigma$}_{1}\!\cdot\!{\bf k}
        \pm\frac{\omega}{2M}\mbox{\boldmath $\sigma$}_{1}\!\cdot\!
                 ({\bf p}'+{\bf p}) \right],            \label{Gam1p}\\
   \bar{u}({\bf p}')\Gamma^{(1)}_{V}u({\bf p}) &=&
      g_{V}\left[ 
      \left\{\left(1+\frac{{\bf p}'\cdot{\bf p}}{4M^2}\right)
 -\frac{i}{4M^2} {\bf p}'\times{\bf p}\cdot\mbox{\boldmath $\sigma$}
      \right\}\ \phi^{0}_{V} \right.\nonumber\\ && \left.
     - \frac{1}{2M}\biggl\{({\bf p}'+{\bf p})
          +i(1+\kappa_{V})\mbox{\boldmath $\sigma$}_{1}\!\times\!{\bf k}
     \biggr\}\!\cdot\!\mbox{\boldmath $\phi$}_{V}\right],    \label{Gam1v}\\
   \bar{u}({\bf p}')\Gamma^{(1)}_{A}u({\bf p}) &=&
      g_{A}\left[ -\frac{1}{2M} \left\{\bm{\sigma}\cdot({\bf p}'+{\bf p}) 
      \right\}\ \phi^{0}_{A} \right.\nonumber\\ && \left.
     + \biggl\{ \bm{\sigma} +\frac{1}{4M^2}(\bm{\sigma}\cdot{\bf p}')\
     \bm{\sigma}\ (\bm{\sigma}\cdot{\bf p})
     \biggr\}\!\cdot\!\mbox{\boldmath $\phi$}_{A}\right],    \label{Gam1a}\\
   \bar{u}({\bf p}')\Gamma^{(1)}_{S}u({\bf p}) &=& g_{S}
 \left[\left(1-\frac{{\bf p}'\cdot{\bf p}}{4M^2}\right)
 -\frac{i}{4M^2} {\bf p}'\times{\bf p}\cdot\mbox{\boldmath $\sigma$}
 \right] ,                                               \label{Gam1s}
\end{eqnarray}
\end{subequations}
\noindent
$\!\!\!$where we defined ${\bf k}={\bf p}'-{\bf p}$ and
$\kappa_{V}=f_{V}/g_{V}$.
In the pseudovector vertex, the upper (lower) sign stands for creation
(absorption) of the pion at the vertex.
In passing we note that the inclusion of the $1/M^2$-terms is necessary
in order to get spin-orbit potentials, like in the case of the OBE-potentials.\\

The complete quark-meson verices are:\\
\noindent (i) Scalar mesons: Including the extra quark-level coupling
\begin{eqnarray}
   \bar{u}({\bf p}')\Gamma_{S}u({\bf p}) &=& g_{S}
 \left(1-\frac{{\bf k}^2}{4m_Q^2}\right)
 \left[\left(1-\frac{{\bf p}'\cdot{\bf p}}{4M^2}\right)
 -\frac{i}{4M^2} {\bf p}'\times{\bf p}\cdot\bm{\sigma} \right],                         
\label{eq:mevert1} \end{eqnarray}

\noindent (ii) Vector mesons:
For the complete vector-meson coupling to the quarks 
\begin{eqnarray*}
&& \Gamma^\mu_V = G_m \gamma^\mu + \frac{1}{{\cal M}} G_e (p'+p)^\mu,\
   G_{m,v} = g_v+f_v\ ,\ G_{e,v} = -f_v\left[1+\frac{k^2}{8m_Q^2} \right],
\end{eqnarray*}
and writing $\Gamma_V = \Gamma_V^{(m)}+\Gamma_V^{(e)}$, 
\begin{subequations}
\label{eq:mevert2}     
\begin{eqnarray}
   \bar{u}({\bf p}')\Gamma^{(m)}_{V}u({\bf p}) &=&
      G_{m,v}\left[ 
      \left\{\left(1+\frac{{\bf p}'\cdot{\bf p}}{4M^2}\right)
 +\frac{i}{4M^2} {\bf p}'\times{\bf p}\cdot\mbox{\boldmath $\sigma$}
      \right\}\ \phi^{0}_{V} \right.\nonumber\\ && \hspace*{1cm} \left.
     + \frac{1}{2M}\biggl\{({\bf p}'+{\bf p})
          +i\mbox{\boldmath $\sigma$}_{1}\!\times\!{\bf k}
     \biggr\}\!\cdot\!\mbox{\boldmath $\phi$}_{V}\right], \\                
   \bar{u}({\bf p}')\Gamma^{(e)}_{V}u({\bf p}) &=&
      G_{e,v}\left[\frac{{\cal E}^\prime+{\cal E}}{{\cal M}}
      \left\{\left(1-\frac{{\bf p}'\cdot{\bf p}}{4M^2}\right)
 -\frac{i}{4M^2} {\bf p}'\times{\bf p}\cdot\mbox{\boldmath $\sigma$}
      \right\}\ \phi^{0}_{V} \right.\nonumber\\ && \hspace*{1cm} \left.
     + \frac{({\bf p}'+{\bf p})}{{\cal M}}
      \left\{\left(1-\frac{{\bf p}'\cdot{\bf p}}{4M^2}\right)
 -\frac{i}{4M^2} {\bf p}'\times{\bf p}\cdot\mbox{\boldmath $\sigma$}
   \right\}  \biggr\}\!\cdot\!\mbox{\boldmath $\phi$}_{V}\right] 
\nonumber\\ &\approx&      
      G_{e,v}\left[2\frac{M}{{\cal M}}
      \left\{\left(1+\frac{{\bf p}^{\prime 2}-{\bf p}'\cdot{\bf p}
 +{\bf p}^2}{4M^2}\right)
 -\frac{i}{4M^2} {\bf p}'\times{\bf p}\cdot\mbox{\boldmath $\sigma$}
      \right\}\ \phi^{0}_{V} 
     + \frac{({\bf p}'+{\bf p})}{{\cal M}}
   \biggr\}\!\cdot\!\mbox{\boldmath $\phi$}_{V}\right] 
\end{eqnarray}
\end{subequations}

\noindent (iii) Axial-vector mesons:
The extra QQ axial-coupling has the vertex 
\begin{eqnarray}
   \bar{u}({\bf p}')\Gamma^{(o)}_{A}u({\bf p}) &=&
      \frac{g_a^\prime}{{\cal M}^2} \left[  \frac{1}{M}\left\{\vphantom{\frac{A}{A}}
 ({\bf p}^\prime\cdot{\bf p}-{\bf p}^2)\bm{\sigma}\cdot{\bf p}' +
 ({\bf p}^\prime\cdot{\bf p}-{\bf p}^{\prime 2})\bm{\sigma}\cdot{\bf p} 
      \right\}\ \phi^{0}_{A} 
 -2i {\bf p}^\prime\times{\bf p}\cdot\!\mbox{\boldmath $\phi$}_{A}\right] 
\nonumber\\ &=& \frac{g_a}{4{\cal M}^2}\left[\frac{1}{M}\left\{
\vphantom{\frac{A}{A}} ({\bf q}\cdot{\bf k}\ \bm{\sigma}\cdot{\bf k}
-{\bf k}^2\ \bm{\sigma}\cdot{\bf q}\right\}\ \phi_A^0 
+2i {\bf q}\times{\bf k}\cdot\bm{\phi}_A\right] \nonumber\\
 &\approx& \frac{g_a}{2{\cal M}^2}\cdot 
 i{\bf q}\times{\bf k}\cdot\bm{\phi}_A,             
\label{Gaxia1} \end{eqnarray}
{\it i.e.} a purely spin-orbit contribution. Using 
\begin{eqnarray*}
&& (\bm{\sigma}\cdot{\bf p}')\ \bm{\sigma}\ (\bm{\sigma}\cdot{\bf p}) = 
   {\bf p}'\ (\bm{\sigma}\cdot{\bf p})  + {\bf p}\ (\bm{\sigma}\cdot{\bf p}')  
 -{\bf p}'\cdot{\bf p}\ \bm{\sigma} -i {\bf p}'\times{\bf p}= \\
&& 2{\bf q} (\bm{\sigma}\cdot{\bf q}) -\frac{1}{2}{\bf k}\ (\bm{\sigma}\cdot{\bf k})
-({\bf q}^2-{\bf k}^2/4)\ \bm{\sigma}+i {\bf q}\times{\bf k}.
\end{eqnarray*}
we obtain for the complete axial-vertex, with ${\cal M} = M$,
\begin{eqnarray}
   \bar{u}({\bf p}')\Gamma_{A}u({\bf p}) &=& 
      g_{A}\left[ -\frac{1}{M} (\bm{\sigma}\cdot{\bf q}) 
      \phi^{0}_{A} 
     + \biggl\{ \bm{\sigma}\left(1-\frac{{\bf q}^2-{\bf k}^2/4}{4M^2}\right)
\right.\nonumber\\ && \left.
+\frac{1}{4M^2}\left(2{\bf q}(\bm{\sigma}\cdot{\bf q})
  -\frac{1}{2}{\bf k}(\bm{\sigma}\cdot{\bf k})\right)
 +\frac{3i}{4M^2} {\bf q}\times{\bf k}             
     \biggr\}\!\cdot\!\mbox{\boldmath $\phi$}_{A}\right]. 
\label{Gaxia2} \end{eqnarray}

\section{One-Boson-Exchange Quark-quark Potentials}
\label{app:OBE}
\subsection{Non-strange Meson-exchange}
\label{app:OBE.a}
For the non-strange mesons the mass differences at the vertices are neglected,
we take at the $QQM$- and $NNM$-vertex the average quark and average
nucleon mass respectively. This implies that we do not include contributions
to the Pauli-invariants $P_7$ and $P_8$.
 For vector-, and diffractive OBE-exchange we  
refer the reader to Ref.~\cite{MRS89}, where the contributions to the different
$\Omega^{(X)}_{i}$'s for baryon-baryon scattering are given in detail.\\
Below, for QQ: $M_n=M_y=M_Q$, and for QN: $M_y=M_Q,M_n=M_N$. For QQ-channels: $g^x_{13}=(g^x_{QQ})_{13}$ and 
$g^x_{24}=(g^x_{QQ})_{24}$ etc., and for QN-channels $g^x_{13}=(g^x_{QQ})_{13}$ and $g^x_{24}=(g^x_{NN})_{24}$ etc.,
where x=p,pv,v,a, and b.
\begin{enumerate}
 \item[(a)]   Pseudoscalar-meson exchange:
      \begin{subequations}
      \begin{eqnarray}
       \Omega^{(P)}_{2a} & = & -g^p_{13}g^p_{24}\left( \frac{{\bf k}^{2}}
           {12M_yM_n} \right) \ \ ,\ \ 
       \Omega^{(P)}_{3a}  =  -g^p_{13}g^p_{24}\left( \frac{1}
           {4M_yM_n}  \right), \label{eq1a} \\
       \Omega^{(P)}_{2b} & = & +g^p_{13}g^p_{24}\left( \frac{{\bf k}^{2}}
           {24M_y^2M_n^2} \right) \ \ ,\ \ 
       \Omega^{(P)}_{3b}  =  +g^p_{13}g^p_{24}\left( \frac{1}
           {8M_y^2M_n^2}  \right), \label{eq1b}.    
         \end{eqnarray}
\hspace{3mm} PV-formulas:
      \begin{eqnarray}
       \Omega^{(P)}_{2a} & = & -f^{pv}_{13}f^{pv}_{24}\left( \frac{{\bf k}^{2}}
           {3 m_{\pi^+}^2} \right) \ \ ,\ \ 
       \Omega^{(P)}_{3a}  =  -f^{pv}_{13}f^{pv}_{24}\left( \frac{1}
           { m_{\pi^+}^2} \right), \label{eq2a} \\
       \Omega^{(P)}_{2b} & = & +f^{pv}_{13}f^{pv}_{24}\left( \frac{{\bf k}^{2}}
           {6 m_{\pi^+}^2 M_y M_n} \right) \ \ ,\ \ 
       \Omega^{(P)}_{3b}  =  +f^{pv}_{13}f^{pv}_{24}\left( \frac{1}
           {2m_{\pi^+}^2 M_y^2M_n^2}  \right), \label{eq2b}.    
         \end{eqnarray}
      \end{subequations}
 \item[(b)]   Vector-meson exchange:
     \begin{eqnarray}  
       \Omega^{(V)}_{1a}&=&
   \left\{g^v_{13}g^v_{24}\left( 1-\frac{{\bf k}^{2}}{2M_yM_n}\right)
           -g^v_{13}f^v_{24}\frac{{\bf k}^{2}}{4{\cal M}M_n} 
      -f^v_{13}g^v_{24}\frac{{\bf k}^{2}}{4{\cal M}M_y}
 \vphantom{\frac{A}{A}}\right. \nonumber\\ && \left. \vphantom{\frac{A}{A}}
           +f^v_{13}f^v_{24}\frac{{\bf k}^{4}}
           {16{\cal M}^{2}M_yM_n}\right\},\ \                 
    \Omega^{(V)}_{1b} =  g^v_{13}g^v_{24}\left(
    \frac{3}{2M_yM_n}\right), \nonumber\\
  \Omega^{(V)}_{2a} &=& -\frac{2}{3} {\bf k}^{2}\,\Omega^{(V)}_{3a}, \ \ 
  \Omega^{(V)}_{2b}  =  -\frac{2}{3} {\bf k}^{2}\,\Omega^{(V)}_{3b}, 
 \nonumber\\
    \Omega^{(V)}_{3a}&=& \left\{
           (g^v_{13}+f^v_{13}\frac{M_y}{{\cal M}})
           (g^v_{24}+f^v_{24}\frac{M_n}{{\cal M}}) 
          -f^v_{13}f^v_{24}\frac{{\bf k}^{2}}{8{\cal M}^{2}} \right\}
            /(4M_yM_n), \nonumber\\                 
    \Omega^{(V)}_{3b}&=& -
           (g^v_{13}+f^v_{13}\frac{M_y}{{\cal M}})
           (g^v_{24}+f^v_{24}\frac{M_n}{{\cal M}}) 
            /(8M_y^2M_n^2), \nonumber\\                 
    \Omega^{(V)}_{4}&=&-\left\{12g^v_{13}g^v_{24}+8(g^v_{13}f^v_{24}+f^v_{13}g^v_{24})
           \frac{\sqrt{M_yM_n}}{{\cal M}} 
     - f^v_{13}f^v_{24}\frac{3{\bf k}^{2}}{{\cal M}^{2}}\right\}
            /(8M_yM_n)              \nonumber\\
       \Omega^{(V)}_{5}&=&- \left\{
           g^v_{13}g^v_{24}+4(g^v_{13}f^v_{24}+f^v_{13}g^v_{24})
           \frac{\sqrt{M_yM_n}}{{\cal M}}  
           +8f^v_{13}f^v_{24}\frac{M_yM_n}{{\cal M}^{2}}\right\}
          /(16M_y^{2}M_n^{2})        \nonumber\\
       \Omega^{(V)}_{6}&=&-\left\{(g^v_{13}g^v_{24}
           +f^v_{13}f^v_{24}\frac{{\bf k}^{2}}{4{\cal M}^{2}})
    \frac{(M_n^{2}-M_y^{2})}{4M_y^{2}M_n^{2}} 
      -(g^v_{13}f^v_{24}-f^v_{13}g^v_{24})
      \frac{1}{\sqrt{{\cal M}^{2}M_yM_n}}\right\}.
 \nonumber\\
 \label{eq2}\end{eqnarray}
 \item[(c)]   Scalar-meson exchange:  \hspace{2em}
      \begin{eqnarray} 
      \Omega^{(S)}_{1a} & = & 
      -g^s_{13} g^s_{24} \left( 1+\frac{{\bf k}^{2}}{4M_yM_n}\right)
       \nonumber\\ &&\nonumber\\
      \Omega^{(S)}_{1b} & = & +g^s_{13} g^s_{24} \left[\frac{1}{2M_yM_n}\right]
  \ \ ,\ \ 
      \Omega^{(S)}_{4}= -g^s_{13} g^s_{24} \left[\frac{1}{2M_yM_n}\right]
       \nonumber\\ &&\nonumber\\
      \Omega^{(S)}_{5} &=& g^s_{13} g^s_{24}
        \left[\frac{1}{16M_y^{2}M_n^{2} }\right] 
  \ \ ,\ \ 
      \Omega^{(S)}_{6}= -g^s_{13} g^s_{24}
        \frac{(M_n^{2}-M_y^{2})}{4M_y^2M_n^2}.
       \label{Eq:scal} \end{eqnarray}
\item[(d)] Axial-vector-exchange $J^{PC}=1^{++}$:
      \begin{eqnarray} 
      \Omega^{(A)}_{2a} & = & -g^a_{13}g^a_{24}\left[
         1-\frac{2{\bf k}^2}{3M_yM_n}\right]
         +\left[\left(g_{13}^A f_{24}^A\frac{M_n}{{\cal M}}
         +f_{13}^A g_{24}^A \frac{M_y}{{\cal M}}\right)
         -f_{13}^A f_{24}^A \frac{{\bf k}^2}{2{\cal M}^2}\right]\
         \frac{{\bf k}^2}{6M_yM_n}
       \nonumber\\ && \nonumber\\
      \Omega^{(A)}_{2b} &=& 
        -g^a_{13}g^a_{24} \left(\frac{3}{2M_yM_n}\right) 
        \nonumber\\ && \nonumber\\
      \Omega^{(A)}_{3}&=&
        -g^a_{13}g^a_{24} \left[\frac{1}{4M_yM_n}\right]
         +\left[\left(g_{13}^A f_{24}^A\frac{M_n}{{\cal M}}
         +f_{13}^A g_{24}^A \frac{M_y}{{\cal M}}\right)
         -f_{13}^A f_{24}^A \frac{{\bf k}^2}{2{\cal M}^2}\right]\
         \frac{1}{2M_yM_n}
       \nonumber\\ && \nonumber\\
	\Omega^{(A)}_{4}  &=&
     -g^a_{13}g^a_{24}   \left[\frac{1}{2M_yM_n}\right] 
      \ \ ,\ \
      \Omega^{(A)}_{6} = 
     -g^a_{13}g^a_{24} \left[\frac{(M_n^{2}-M_y^{2})}{4M_y^2M_n^2}\right]
       \nonumber\\ && \nonumber\\
      \Omega^{(A)'}_{5} & = &
     -g^a_{13}g^a_{24}   \left[\frac{2}{M_yM_n}\right] 
         \label{eq:axi1} \end{eqnarray}
Here, we used the B-field description with $\alpha_r=1$, 
see \cite{RNY10a} Appendix A.              
 The detailed treatment of the potential proportional to $P_5'$, i.e. 
 with $\Omega_5^{(A)'}$, is given in \cite{RNY10a}, Appendix~B.
\item[(e)] Axial-vector mesons with $J^{PC}=1^{+-}$: 
      \begin{eqnarray} 
       \Omega^{(B)}_{2a} & = & +f^B_{13}f^B_{24}\frac{(M_n+M_y)^2}{m_B^2}
       \left(1-\frac{{\bf k}^2}{4M_yM_n}\right)
       \left( \frac{{\bf k}^{2}}{12M_yM_n} \right),\ \ 
       \Omega^{(B)}_{2b}   =   +f^B_{13}f^B_{24}\frac{(M_n+M_y)^2}{m_B^2}
       \left( \frac{{\bf k}^{2}}{8M_y^2M_n^2} \right)     
     \nonumber\\ 
       \Omega^{(B)}_{3a} & = & +f^B_{13}f^B_{24}\frac{(M_n+M_y)^2}{m_B^2}
       \left(1-\frac{{\bf k}^2}{4M_yM_n}\right)
       \left( \frac{1}{4M_yM_n} \right),\ \             
       \Omega^{(B)}_{3b}   =   +f^B_{13}f^B_{24}\frac{(M_n+M_y)^2}{m_B^2}
       \left( \frac{3}{8M_y^2M_n^2} \right). \nonumber\\     
     \label{eq:bxi1} \end{eqnarray}
 \item[(f)]   Diffractive-exchange (pomeron, $f, f', A_{2}$): \\
         The $\Omega^{D}_{i}$ are the same as for scalar-meson-exchange
         Eq.(\ref{Eq:scal}), but with
         $\pm g_{13}^{S}g_{24}^{S}$ replaced by
         $\mp g_{13}^{D}g_{24}^{D}$, and except for the zero in the form factor.
\item[(g)] Odderon-exchange:              
         The $\Omega^{O}_{i}$ are the same as for vector-meson-exchange
         Eq.(ref{eq2}), but with
         $ g_{13}^{V}\rightarrow g_{13}^{O}$, 
         $ f_{13}^{V}\rightarrow f_{13}^{O}$ and similarly for the couplings
         with the 24-subscript.

\end{enumerate}

As in Ref.~\cite{MRS89} in the derivation of the expressions for $\Omega_i^{(X)}$, 
given above, $M_y$ and $M_n$ denote the mean hyperon and nucleon
mass, respectively \begin{math} M_y=(M_{1}+M_{3})/2 \end{math}
and \begin{math} M_n=(M_{2}+M_{4})/2 \end{math},
 and $m$ denotes the mass of the exchanged meson.
Moreover, the approximation                            
        \begin{math}
              1/ M^{2}_{N}+1/ M^{2}_{Y}\approx
              2/ M_nM_y,
        \end{math}
is used, which is rather good since the mass differences
between the baryons are not large.\\

\subsection{One-Boson-Exchange Interactions in Configuration Space I}
\label{app.IIIb}
In configuration space the BB-interactions are described by potentials
of the general form
\begin{subequations}
\begin{eqnarray}
 V &=& \left\{\vphantom{\frac{A}{A}} V_C(r) + V_\sigma(r)
\mbox{\boldmath $\sigma$}_1\cdot\mbox{\boldmath $\sigma$}_2
 + V_T(r) S_{12} + V_{SO}(r) {\bf L}\cdot{\bf S} + V_Q(r) Q_{12}
 \right.\nonumber\\ && \left.
 + V_{ASO}(r)\ \frac{1}{2}(\mbox{\boldmath $\sigma$}_1-
  \mbox{\boldmath $\sigma$}_2)\cdot{\bf L}
 -\frac{1}{2M_yM_n}\left(\vphantom{\frac{A}{A}} 
\mbox{\boldmath $\nabla$}^2 V^{n.l.}(r) + V^{n.l.}(r) 
 \mbox{\boldmath $\nabla$}^2\right)
\right\}\cdot {\cal P}, \\
 V^{n.l.} &=& \left\{\vphantom{\frac{A}{A}} \varphi_C(r) + \varphi_\sigma(r)
\mbox{\boldmath $\sigma$}_1\cdot\mbox{\boldmath $\sigma$}_2
 + \varphi_T(r) S_{12}\right\}\cdot {\cal P}, 
 \label{eq:3b.a}\end{eqnarray}
\end{subequations}
where for non-strange mesons ${\cal P}=1$, and
\begin{subequations}
\begin{eqnarray}
 S_{12} &=& 3 (\mbox{\boldmath $\sigma$}_1\cdot\hat{r})
 (\mbox{\boldmath $\sigma$}_2\cdot\hat{r}) -
 (\mbox{\boldmath $\sigma$}_1\cdot\mbox{\boldmath $\sigma$}_2), \\
 Q_{12} &=& \frac{1}{2}\left[\vphantom{\frac{A}{A}} 
 (\mbox{\boldmath $\sigma$}_1\cdot{\bf L})(\mbox{\boldmath $\sigma$}_2\cdot{\bf L})
 +(\mbox{\boldmath $\sigma$}_2\cdot{\bf L})(\mbox{\boldmath $\sigma$}_1\cdot{\bf L})
 \right], \\
 \phi(r) &=& \phi_C(r) + \phi_\sigma(r) 
 \mbox{\boldmath $\sigma$}_1\cdot\mbox{\boldmath $\sigma$}_2, 
 \label{eq:3b.b}\end{eqnarray}
\end{subequations}
For the basic functions for the Fourier transforms with gaussian form factors,
we refer to Refs.~\cite{NRS78,MRS89}.                           
For the details of the Fourier transform for the potentials with $P_5'$, which 
occur in the case of the axial-vector mesons with $J^{PC}=1^{++}$, we refer 
to Ref.~\cite{RNY10a} Appendix B. 

\noindent (a)\ Pseudoscalar-meson-exchange:
\begin{subequations}
\begin{eqnarray}
  V_{PS}(r) &=& \frac{m}{4\pi}\left[ g^p_{13}g^p_{24}\frac{m^2}{4M_yM_n}
 \left(\frac{1}{3}(\mbox{\boldmath $\sigma$}_1\cdot\mbox{\boldmath $\sigma$}_2)\
 \phi_C^1 + S_{12} \phi_T^0\right)\right] {\cal P}, \\
  V_{PS}^{n.l.}(r) &=& \frac{m}{4\pi}\left[ g^p_{13}g^p_{24}\frac{m^2}{4M_yM_n}
 \left(\frac{1}{3}(\mbox{\boldmath $\sigma$}_1\cdot\mbox{\boldmath $\sigma$}_2)\
 \phi_C^1 + S_{12} \phi_T^0\right)\right] {\cal P}. 
 \label{eq:3b.1}\end{eqnarray}
\end{subequations}
\noindent (b)\ Vector-meson-exchange:          
\begin{subequations}
\begin{eqnarray}
&& V_{V}(r) = \frac{m}{4\pi}\left[\left\{ g^v_{13}g^v_{24}\left[ \phi_C^0 +
 \frac{m^2}{2M_yM_n} \phi_C^1 
\right]
\right.\right.\nonumber\\ && \left.\left.  \hspace{0cm} 
 +\left[g^v_{13}f^v_{24}\frac{m^2}{4{\cal M}M_n}
 +f^v_{13}g^v_{24}\frac{m^2}{4{\cal M}M_y}\right] \phi_C^1 + f^v_{13}f^v_{24}
\frac{m^4}{16{\cal M}^2 M_y M_n} \phi_C^2\right\}
  \right.\nonumber\\ && \left.  \hspace{0cm} 
 +\frac{m^2}{6M_yM_n}\left\{\left[ \left(g^v_{13}+f^v_{13}\frac{M_y}{{\cal M}}\right)\cdot
 \left(g^v_{24}+f^v_{24}\frac{M_n}{{\cal M}}\right)\right] \phi_C^1 
 +f^v_{13} f^v_{24}\frac{m^2}{8{\cal M}^2} \phi_C^2\right\}
 (\mbox{\boldmath $\sigma$}_1\cdot\mbox{\boldmath $\sigma$}_2)\
  \right.\nonumber\\ && \left.  \hspace{0cm} 
 -\frac{m^2}{4M_yM_n}\left\{\left[ \left(g^v_{13}+f^v_{13}\frac{M_y}{{\cal M}}\right)\cdot
 \left(g^v_{24}+f^v_{24}\frac{M_n}{{\cal M}}\right)\right] \phi_T^0 
 +f^v_{13} f^v_{24}\frac{m^2}{8{\cal M}^2} \phi_T^1\right\} S_{12}
  \right.\nonumber\\ && \left.  \hspace{0cm} 
 -\frac{m^2}{M_yM_n}\left\{\left[ \frac{3}{2}g^v_{13}g^v_{24}
 +\left(g^v_{13}f^v_{24}+f^v_{13}g^v_{24}\right)
 \frac{\sqrt{M_yM_n}}{{\cal M}}\right] \phi_{SO}^0 
 +\frac{3}{8}f^v_{13} f^v_{24}\frac{m^2}{{\cal M}^2} \phi_{SO}^1\right\} {\bf L}\cdot{\bf S}
  \right.\nonumber\\ && \left.  \hspace{0cm} 
 +\frac{m^4}{16M_y^2M_n^2}\left\{\left[ g^v_{13}g^v_{24}
 +4\left(g^v_{13}f^v_{24}+f^v_{13}g^v_{24}\right)
 \frac{\sqrt{M_yM_n}}{{\cal M}} 
 +8f^v_{13}f^v_{24}\frac{M_yM_n}{{\cal M}^2}\right]\right\} 
  \cdot\right.\nonumber\\ && \left.  \hspace{0cm} \times
\frac{3}{(mr)^2} \phi_T^0 Q_{12}
 -\frac{m^2}{M_yM_n}\left\{\left[ 
 \left(g^v_{13}g^v_{24}-f^v_{13}f^v_{24}\frac{m^2}{{\cal M}^2}\right)
  \frac{(M_n^2-M_y^2)}{4M_yM_n}
  \right.\right.\right.\nonumber\\ && \left.\left.\left.  \hspace{0cm} 
  -\left(g^v_{13}f^v_{24}-f^v_{13}g^v_{24}\right)\frac{\sqrt{M_yM_n}}{{\cal M}}\right] \phi_{SO}^0
 \right\}\cdot\frac{1}{2}\left(
 \mbox{\boldmath $\sigma$}_1-\mbox{\boldmath $\sigma$}_2\right)\cdot{\bf L}\right] {\cal P},
\\
&& V_{V}^{n.l.}(r) = \frac{m}{4\pi}\left[ \frac{3}{2} g^v_{13}g^v_{24}\ \phi_C^0 
  \right.\nonumber\\ && \left.  \hspace{0cm} 
 +\frac{m^2}{6M_yM_n}\left\{\left[ \left(g^v_{13}+f^v_{13}\frac{M_y}{{\cal M}}\right)\cdot
 \left(g^v_{24}+f^v_{24}\frac{M_n}{{\cal M}}\right)\right] \phi_C^1 \right\}
 (\mbox{\boldmath $\sigma$}_1\cdot\mbox{\boldmath $\sigma$}_2)\
  \right.\nonumber\\ && \left.  \hspace{0cm} 
 -\frac{m^2}{4M_yM_n}\left\{\left[ \left(g^v_{13}+f^v_{13}\frac{M_y}{{\cal M}}\right)\cdot
 \left(g^v_{24}+f^v_{24}\frac{M_n}{{\cal M}}\right)\right] \phi_T^0 \right\} S_{12}
\right] {\cal P}. 
 \label{eq:3b.2}\end{eqnarray}
\end{subequations}
Note: the non-local tensor and "associated" spin-spin terms are 
not included in ESC16 model.\\

\noindent (c)\ Scalar-meson-exchange:          
\begin{eqnarray}
 V_{S}(r) &=& -\frac{m}{4\pi}\left[ g^s_{13}g^s_{24}\left\{\left[ \phi_C^0 
 -\frac{m^2}{4M_yM_n} \phi_C^1\right] + \frac{m^2}{2M_yM_n} \phi_{SO}^0\ {\bf L}\cdot{\bf S}
 +\frac{m^4}{16M_y^2M_n^2}
 \cdot\right.\right.\nonumber\\ && \left.\left. \times
\frac{3}{(mr)^2} \phi_T^0 Q_{12} 
 +\frac{m^2}{M_yM_n} \left[\frac{(M_n^2-M_y^2)}{4M_yM_n}\right] \phi_{SO}^0\cdot
 \frac{1}{2}\left(\mbox{\boldmath $\sigma$}_1-\mbox{\boldmath $\sigma$}_2\right)\cdot{\bf L}
  \right.\right.\nonumber\\ && \left.\left.  \hspace{0.0cm}
  +\frac{1}{4M_yM_n}\left(\mbox{\boldmath $\nabla$}^2 \phi_C^0 
 + \phi_C^0 \mbox{\boldmath $\nabla$}^2\right) \right\}\right] {\cal P}.
 \label{eq:3b.3}\end{eqnarray}
\noindent (d)\ Axial-vector-meson exchange $J^{PC}=1^{++}$:
\begin{eqnarray}
&& V_{A}(r) = -\frac{m}{4\pi}\left[ 
 \left\{ g^a_{13}g^a_{24}\left(\phi_C^0 +\frac{2m^2}{3M_yM_n} \phi_C^1\right)
 +\frac{m^2}{6M_yM_n}\left(g^a_{13}f^a_{24}\frac{M_n}{{\cal M}}
 +f^a_{13}g^a_{24}\frac{M_y}{{\cal M}}\right)\phi_C^1
\right.\right.\nonumber\\ && \left.\left.
 +f^a_{13}f^a_{24}\frac{m^4}{12M_yM_n{\cal M}^2}\phi_C^2\right\}
 (\mbox{\boldmath $\sigma$}_1\cdot\mbox{\boldmath $\sigma$}_2)
  -\frac{3}{4M_yM_n} g^a_{13}g^a_{24}\left(\mbox{\boldmath $\nabla$}^2 \phi_C^0 
 + \phi_C^0 \mbox{\boldmath $\nabla$}^2\right) 
 (\mbox{\boldmath $\sigma$}_1\cdot\mbox{\boldmath $\sigma$}_2)
 \right.\nonumber\\ && \left. 
 - \frac{m^2}{4M_yM_n}\left\{\left[g^a_{13}g^a_{24}-2\left(g^a_{13}f^a_{24}
 \frac{M_n}{{\cal M}}+f^a_{13}g^a_{24}\frac{M_y}{{\cal M}}\right)\right] \phi_T^0
 -f^a_{13}f^a_{24}\frac{m^2}{{\cal M}^2} \phi_T^1\right\} S_{12}
 \right.\nonumber\\ && \left. 
 +\frac{m^2}{2M_yM_n}g^a_{13}g^a_{24} \left\{\phi_{SO}^0\ {\bf L}\cdot{\bf S}
 +\frac{m^2}{M_yM_n} \left[\frac{(M_n^2-M_y^2)}{4M_yM_n}\right] \phi_{SO}^0\cdot
 \frac{1}{2}\left(\mbox{\boldmath $\sigma$}_1-\mbox{\boldmath $\sigma$}_2\right)\cdot{\bf L}
 \right\}\right] {\cal P}.     
 \label{eq:3b.4}\end{eqnarray}
\noindent (e)\ Axial-vector-meson exchange $J^{PC}=1^{+-}$:
\begin{subequations}
\begin{eqnarray}
 V_{B}(r) &=& -\frac{m}{4\pi}\frac{(M_n+M_y)^2}{m^2}\left[ 
 f^B_{13}f^B_{24}\left\{\frac{m^2}{12M_yM_n}\left(\phi_C^1+
 \frac{m^2}{4M_yM_n} \phi_C^2\right)
 (\mbox{\boldmath $\sigma$}_1\cdot\mbox{\boldmath $\sigma$}_2)
 \right.\right.\nonumber\\ && \left.\left. 
  -\frac{m^2}{8M_yM_n}\left(\mbox{\boldmath $\nabla$}^2 \phi_C^1 
 + \phi_C^1 \mbox{\boldmath $\nabla$}^2\right) 
 (\mbox{\boldmath $\sigma$}_1\cdot\mbox{\boldmath $\sigma$}_2)
 +\left[\frac{m^2}{4M_yM_n}\right] \phi^0_T\ S_{12}\right\}\right] {\cal P}, \\
 V_{B}^{n.l.}(r) &=& -\frac{m}{4\pi}\frac{(M_n+M_y)^2}{m^2}\left[ 
 f^B_{13}f^B_{24}\left\{
  \frac{3m^2}{4M_yM_n} \left(\frac{1}{3}  
  \mbox{\boldmath $\sigma$}_1\cdot\mbox{\boldmath $\sigma$}_2\ \phi_C^1
  + S_{12}\ \phi_T^0\right)\right\}\right] {\cal P}. 
 \label{eq:3b.5}\end{eqnarray}
\end{subequations}
\noindent (f)\ Diffractive exchange:           
\begin{eqnarray}
&& V_{D}(r) = \frac{m_P}{4\pi}\left[ g^D_{13}g^D_{24} 
 \frac{4}{\sqrt{\pi}}\frac{m_P^2}{{\cal M}^2}\cdot\left[\left\{
 1+\frac{m_P^2}{2M_yM_n}(3-2 m_P^2r^2) + \frac{m_P^2}{M_yM_n} {\bf L}\cdot{\bf S}
  \right.\right.\right.\nonumber\\ && \left.\left.\left.  \hspace{0.75cm} 
 +\left(\frac{m_P^2}{2M_yM_n}\right)^2 Q_{12}
 +\frac{m_P^2}{M_yM_n} \left[\frac{(M_n^2-M_y^2)}{4M_yM_n}\right]\cdot
 \frac{1}{2}\left(\mbox{\boldmath $\sigma$}_1-\mbox{\boldmath $\sigma$}_2\right)\cdot{\bf L}
 \right\}\ e^{-m_P^2r^2} 
  \right.\right.\nonumber\\ && \left.\left.  \hspace{1.5cm} 
  +\frac{1}{4M_yM_n}\left(\mbox{\boldmath $\nabla$}^2 e^{-m_P^2r^2} 
 + e^{-m_P^2r^2}\mbox{\boldmath $\nabla$}^2\right) \right]\right] {\cal P}.
 \label{eq:3b.6}\end{eqnarray}
\noindent (g)\ Odderon-exchange:                      
\begin{subequations}
\begin{eqnarray}
 V_{O,C}(r) &=& +\frac{g^O_{13}g^O_{24}}{4\pi}\frac{8}{\sqrt{\pi}}\frac{m_O^5}{{\cal M}^4}
 \left[\left(3-2m_O^2r^2\right) \right.\nonumber\\ && \left. 
 -\frac{m_O^2}{M_yM_n}\left( 15 - 20 m_O^2r^2+4 m_O^4r^4\right)
 \right]\exp(-m_O^2 r^2)\ , \\
 V_{O,n.l.}(r) &=& -\frac{g^O_{13}g^O_{24}}{4\pi}\frac{8}{\sqrt{\pi}}\frac{m_O^5}{{\cal M}^4}
 \frac{3}{4M_yM_n}\left\{\mbox{\boldmath $\nabla$}^2
 \left[(3-2m_O^2r^2)\exp(-m_O^2 r^2)\right]+ \right.\nonumber\\
 && \left. + \left[(3-2m_O^2r^2)\exp(-m_O^2 r^2)\right] 
 \mbox{\boldmath $\nabla$}^2 \right\}\ , \\
 V_{O,\sigma}(r) &=& 
-\frac{g^O_{13}g^O_{24}}{4\pi}\frac{8}{3\sqrt{\pi}}\frac{m_O^5}{{\cal M}^4}
 \frac{m_O^2}{M_yM_n}
 \left[15-20 m_O^2r^2+4 m_O^4 r^4\right]\exp(-m_O^2 r^2)\cdot \nonumber\\
 && \times\left(1+\kappa^O_{13}\frac{M_y}{\cal M}\right) 
 \left(1+\kappa^O_{24}\frac{M_n}{\cal M}\right) 
 , \\
 V_{O,T}(r) &=& -\frac{g^O_{13}g^O_{24}}{4\pi}\frac{8}{3\sqrt{\pi}}\frac{m_O^5}{{\cal M}^4}
 \frac{m_O^2}{M_yM_n}\cdot m_O^2 r^2
 \left[7-2 m_O^2r^2\right]\exp(-m_O^2 r^2)\cdot \nonumber\\
 && \times\left(1+\kappa^O_{13}\frac{M_y}{\cal M}\right) 
 \left(1+\kappa^O_{24}\frac{M_n}{\cal M}\right) 
 , \\
 V_{O,SO}(r) &=& -\frac{g^O_{13}g^O_{24}}{4\pi}\frac{8}{\sqrt{\pi}}\frac{m_O^5}{{\cal M}^4}
 \frac{m_O^2}{M_yM_n} \left[5-2 m_O^2r^2\right]\exp(-m_O^2 r^2)\cdot \nonumber\\
 && \times\left\{3+\left(\kappa^O_{13}+\kappa^O_{24}\right)\frac{\sqrt{M_yM_n}}{\cal M}\right\}
 , \\
 V_{O,Q}(r) &=& +\frac{g^O_{13}g^O_{24}}{4\pi}\frac{2}{\sqrt{\pi}}\frac{m_O^5}{{\cal M}^4}
 \frac{m_O^4}{M_y^2M_n^2} 
 \left[7-2 m_O^2r^2\right]\exp(-m_O^2 r^2)\cdot \nonumber\\
 && \times\left\{1+4\left(\kappa^O_{13}+\kappa^O_{24}\right)\frac{\sqrt{M_yM_n}}{\cal M}
 +8\kappa_{13}\kappa_{24}\frac{M_yM_n}{{\cal M}^2}\right\}
 , \\
 V_{O,ASO}(r) &=& -\frac{g^O_{13}g^O_{24}}{4\pi}\frac{4}{\sqrt{\pi}}\frac{m_O^5}{{\cal M}^4}
 \frac{m_O^2}{M_yM_n} \left[5-2 m_O^2r^2\right]\exp(-m_O^2 r^2)\cdot \nonumber\\
 &&  \times\left\{ \frac{M_n^2-M_y^2}{M_yM_n}
-4\left(\kappa^O_{24}-\kappa^O_{13}\right)
 \frac{\sqrt{M_yM_n}}{\cal M} \right\}.
 \label{eq:3b.7}\end{eqnarray}
\end{subequations}

\subsection{Strange Meson-exchange}
\label{app:OBE.d}
The rules for hypercharge nonzero exchange have been given in 
Ref.~\cite{NRS77}, see also \cite{NRY19b}. 
The potentials for non-zero hypercharge exchange ($K, K^*,\kappa, K_A, K_B)$ 
are obtained from the expressions   
given in the previous subsections for non-strange mesons by taking
care of the following points: 
(a) For strange meson exchange ${\cal P}=-{\cal P}_x {\cal P}_\sigma$.       
(b) In the latter case one has to replace both $M_n$ and $M_y$ by 
$\sqrt{M_yM_n}$, and reverse the sign of the antisymmetric spin orbit.
\begin{flushleft}
\rule{16cm}{0.5mm}
\end{flushleft}
\section{Additional One-Boson-Exchange QQ-Potentials}
\label{app:OBE2}
The extra vertices at the quark-level generate additional OBE-potentials.  
In the case of the vector mesons the extra vertex gives a change in the couplings
\begin{eqnarray*}
&& g_v \rightarrow g_v'= g_v-f_v\frac{{\bf k}^2}{4{\cal M}m_Q},\ 
 f_v \rightarrow f_v'= f_v-f_v\frac{{\bf k}^2}{4m_Q^2},\
   g_s \rightarrow g_s + g_s\frac{{\bf k}^2}{4m_Q^2}.
 \label{extra.1}\end{eqnarray*}
The extra vertices at the quark-level generate additional OBE-potentials.  
Neglecting the ${\bf k}^4$ etc terms we obtain the following contributions:
\begin{enumerate}
 \item[(a)]   Pseudoscalar-meson exchange: no additional potentials.
 \item[(b)]   Vector-meson exchange:
     \begin{eqnarray}  
       \Delta\Omega^{(V)}_{1a}&=&
   -\bigl\{g^v_{13}f^v_{24}+f^v_{13}g^v_{24}\bigr]\ 
    \frac{{\bf k}^2}{4{\cal M}m_Q},\ \ 
    \Delta\Omega^{(V)}_{1b} =  0, \nonumber\\          
  \Delta\Omega^{(V)}_{2a} &=& -\frac{2}{3} {\bf k}^{2}\,\Delta\Omega^{(V)}_{3a}=0, \ \ 
  \Delta\Omega^{(V)}_{2b}  =  -\frac{2}{3} {\bf k}^{2}\,\Delta\Omega^{(V)}_{3b}=0, 
 \nonumber\\
    \Delta\Omega^{(V)}_{3a}&=& -\left\{
            (g^v_{13}+f^v_{13}\frac{M_y}{{\cal M}})\
           f^v_{24}\left(1+\frac{M_y}{m_Q}\right)
           +(g^v_{24}+f^v_{24}\frac{M_n}{{\cal M}})\
           f^v_{13}\left(1+\frac{M_n}{m_Q}\right) \right\}\
            \frac{{\bf k}^2}{4{\cal M} m_Q}/(4M_yM_n), \nonumber\\                 
    \Delta\Omega^{(V)}_{4}&=& +\biggl\{
     \left(3+2\frac{\sqrt{M_yM_n}}{m_Q}\right)(g^v_{13}f^v_{24}+f^v_{13}g^v_{24})
     +4f^v_{13}f^v_{24}\frac{\sqrt{M_yM_n}}{{\cal M}}\biggr\}
     \left(\frac{{\bf k}^2}{4{\cal M}m_Q}\right)/(2M_yM_n), 
                        \nonumber\\
    \Delta\Omega^{(V)}_{5}&=& +\biggl\{
     \left(1+4\frac{\sqrt{M_yM_n}}{m_Q}\right)(g^v_{13}f^v_{24}+f^v_{13}g^v_{24})
     +8f^v_{13}f^v_{24}\frac{\sqrt{M_yM_n}}{{\cal M}}\biggr\}
     \left(\frac{{\bf k}^2}{4{\cal M}m_Q}\right)/(16M_y^2M_n^2), 
                        \nonumber\\
       \Delta\Omega^{(V)}_{6}&=& 0.                      
 \label{obe2.1}\end{eqnarray}
 \item[(c)]   Scalar-meson exchange:  \hspace{2em}
      \begin{eqnarray} 
      \Delta\Omega^{(S)}_{1a} & = & 
      -g^s_{13} g^s_{24}\ \frac{{\bf k}^2}{2m_Q^2}\ ,\        
      \Delta\Omega^{(S)}_{1b}  = 0,                                    
       \nonumber\\ &&\nonumber\\
      \Delta\Omega^{(S)}_{4} &=& -g^s_{13} g^s_{24}\ 
       \frac{{\bf k}^2}{4m_Q^2}\left[\frac{1}{M_yM_n}\right]\ ,\
      \Delta\Omega^{(S)}_{5} = g^s_{13} g^s_{24}\
        \frac{{\bf k}^2}{4m_Q^2}\left[\frac{1}{8M_y^2M_n^2}\right], 
       \nonumber\\ &&\nonumber\\
      \Delta\Omega^{(S)}_{6} &=& -g^s_{13} g^s_{24}\ 
        \frac{(M_n^{2}-M_y^{2})}{4M_y^2M_n^2}\ \frac{{\bf k}^2}{2m_Q^2}.
 \label{obe2.2}\end{eqnarray}
item[(d)]   See below.                                
\end{enumerate}
The transcription to configuration space potentials of these additional 
Pauli-invariants is similar to that in section~\ref{app:OBE} 
and is readily done. The results are
\begin{enumerate}
	\item[(a)]   Pseudoscalar-meson exchange: $\Delta V_P(r)=0$.        
 \item[(b)]   Vector-meson exchange:
     \begin{eqnarray}  
	     \Delta V_V(r) &=& \frac{m}{4\pi}\ \frac{m^2}{4{\cal M} m_Q}\
	     \biggl[\bigl(g^v_{13}f^v_{24}+f^v_{13}g^v_{24}\bigr)\ \phi_C^1(r)
	     \nonumber\\ && -\frac{m^2}{6M_yM_n}\ \left\{
	     \left(g^v_{13}+f^v_{13}\frac{M_y}{{\cal M}}\right)\
           f^v_{24}\left(1+\frac{M_y}{m_Q}\right)
           +\left(g^v_{24}+f^v_{24}\frac{M_n}{{\cal M}}\right)\
	     f^v_{13}\left(1+\frac{M_n}{m_Q}\right) \right\}\ \phi_C^2(r)\ 
     (\bm{\sigma}_1\cdot\bm{\sigma}_2) 
	     \nonumber\\ && -\frac{m^2}{4M_yM_n}\ \left\{
	     \left(g^v_{13}+f^v_{13}\frac{M_y}{{\cal M}}\right)\
           f^v_{24}\left(1+\frac{M_y}{m_Q}\right)
           +\left(g^v_{24}+f^v_{24}\frac{M_n}{{\cal M}}\right)\
	     f^v_{13}\left(1+\frac{M_n}{m_Q}\right) \right\}\ \phi_T^1(r)\ S_{12}
     \nonumber\\ && -\frac{m^2}{2M_yM_n}\ \left\{
     \left(3+2\frac{\sqrt{M_yM_n}}{m_Q}\right)(g^v_{13}f^v_{24}+f^v_{13}g^v_{24})
	     +4f^v_{13}f^v_{24}\frac{\sqrt{M_yM_n}}{{\cal M}}\right\}\ \phi_{SO}^1(r)\
     {\bf L}\cdot{\bf S} 
     \nonumber\\ && +\frac{m^4}{16M^2_yM^2_n}\ \left\{
     \left(1+4\frac{\sqrt{M_yM_n}}{m_Q}\right)
	     \left(g^v_{13}f^v_{24}+f^v_{13}g^v_{24}\right)
	     +8f^v_{13}f^v_{24}\frac{\sqrt{M_yM_n}}{{\cal M}}\right\} 
     \frac{3}{(mr)^2}\ \phi_T^1 Q_{12} \biggr].
 \label{obe2.3}\end{eqnarray}
 \item[(b)]   Scalar-meson exchange:
  \begin{eqnarray}  
  \Delta V_S(r) &=& -\frac{g_{13}^sg_{24}^s}{4\pi}\ \frac{m^3}{2 m_Q^2}\ \biggl[
	  \phi_C^1(r) +\frac{m^2}{2M_yM_n}\ \phi_{SO}^1(r)\ {\bf L}\cdot{\bf S}
	   +\frac{m^4}{16M_y^2M_n^2}
	  \frac{3}{(mr)^2}\ \phi_T^1\ Q_{12} \nonumber\\ && 
 +\frac{m^2}{M_yM_n} \left[\frac{(M_n^2-M_y^2)}{4M_yM_n}\right] \phi_{SO}^1\cdot
	  \frac{1}{2}\left(\mbox{\boldmath $\sigma$}_1-\mbox{\boldmath $\sigma$}_2\right)\cdot{\bf L}
	  \biggr].
 \label{obe2.4}\end{eqnarray}
 \item[(d)]   Axial-vector-meson exchange:
The additional vertex for the axial-vector mesons is
\begin{eqnarray}
\Delta\Gamma_\mu(p',p;k) &=& \frac{ig_a'}{{\cal M}^2}
\varepsilon^{\mu\kappa\alpha\beta} \gamma_\kappa p_\alpha' p_\beta 
 \label{obe2.11}\end{eqnarray}
where we take ${\cal M}=M_N$.
Restriction to terms of order $1/M_N^2$ the extra axial-vector exchange 
potential becomes
\begin{eqnarray}
\Delta V_A(p',p) &=& \frac{ig_a g_a'}{{\cal M}^2} \biggl[
[\bar{u}(p_1',s_1')u(p_1,s_1)]\ 
[\bar{u}(p_2',s_2')\gamma^\mu\gamma_5 u(p_2,s_2)]\ 
\varepsilon_{\mu\kappa\alpha\beta}\gamma_1^\kappa p_1^{\prime\alpha}p_1^\beta
\nonumber\\ && +
[\bar{u}(p_1',s_1') \gamma^\mu\gamma_5u(p_1,s_1)]\ 
[\bar{u}(p_2',s_2') u(p_2,s_2)]\ 
\varepsilon_{\mu\kappa\alpha\beta}\gamma_2^\kappa p_2^{\prime\alpha}p_2^\beta
 \biggr]\ \left({\bf k}^2+m_A^2\right)^{-1}.
 \label{obe2.12}\end{eqnarray}
 With Pauli-spinors this gives
\begin{eqnarray}
\Delta V_A(p',p) &=& -8\frac{g_a g_a'}{{\cal M}^2} \biggl[
	\frac{i}{2}(\bm{\sigma}_1+\bm{\sigma}_2)\cdot{\bf q}\times{\bf k}
 -\frac{1}{4}\left( \bm{\sigma}_1\cdot{\bf k}\ \bm{\sigma}_2\cdot{\bf k}      
	-\frac{1}{3}{\bf k}^2\ \bm{\sigma}_1\cdot\bm{\sigma}_2\right)
	\nonumber\\ && 
	-\frac{1}{6}{\bf k}^2\ \bm{\sigma}_1\cdot\bm{\sigma}_2 \biggr]\
\left({\bf k}^2+ m_A^2\right)^{-1}. 
 \label{obe2.13}\end{eqnarray}
 In configuration space this leads to 
 \begin{eqnarray}
 \Delta V_A(r) &=& - g_ag'_a\ \frac{m_A}{4\pi}  \frac{m_A^2}{{\cal M}^2}\biggl[
 8 \phi_{SO}^0(r)\ {\bf L}\cdot{\bf S} + 2\phi_T^0(r)\ S_{12} 
	 +\frac{4}{3} \phi_C^1(r)\ (\bm{\sigma}_1\cdot\bm{\sigma}_2) \biggr]
 \label{obe2.14}\end{eqnarray}
 {\it In order that folding with the nucleon wave function agrees in the 
 1/M expansion $g_a'=g_a$.}
\end{enumerate}
\noindent {\bf Remark:} For the Quark-Nucleon additional potentials
there is a factor 1/2. This because there is only an additional
coupling in only one of the vertices of the Feynman-diagram.
\begin{flushleft}
\rule{16cm}{0.5mm}
\end{flushleft}
\section{Quarks and Meson-pairs}                
\label{app:MMQQ}        
\begin{figure}   
 \begin{center}
 \resizebox{7.25cm}{!}        
 {\includegraphics[205,675][405,875]{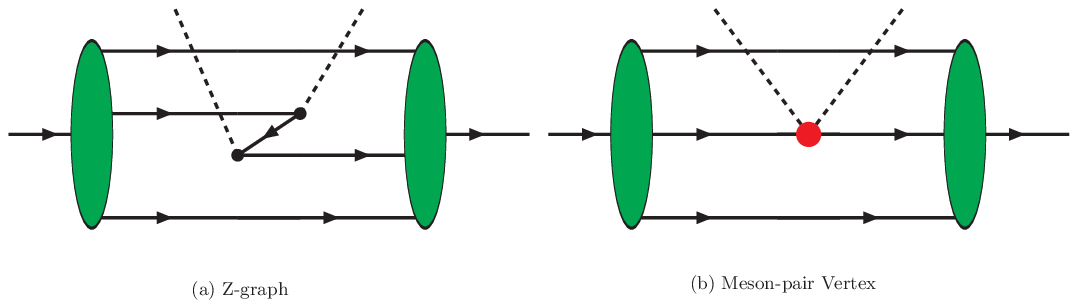}} 
 \caption{Negative-energy quark contribution $\Rightarrow$ MMQQ-coupling}       
 \label{fig.mpeq}
 \end{center}
 \end{figure}

In the Nijmegen models it was in general assumed that negative-energy
nucleons and hyperons are suppressed at low energies and nuclear
densities. In ESC-models it is assumed that in principle the effects
of the negative-energy baryons (Z-graph's) are eventually included effectively 
in the meson-pair couplings to the baryons. 
The same is assumed for the internal quark negative-energy states. 
This is illustrated in Fig.~\ref{fig.mpeq}: the 
Z-graph (a) is included into the meson-meson-quark-quark (MMQQ) coupling.
Then, assuming that the negative-energy
contributions from the baryons are negligible we can suppose that the 
complete MPE in the baryonic nuclear force can be generated
by relating the meson-pair coupling to the quarks from that to the
baryons, similarly as is done in this paper for the meson-couplings
to the quarks. 

\begin{thebibliography}{99}
\bibitem{dercopax}       
Note that in this paper we suppose that $f_A$ does not contain the 
one-pion-pole etc. In momentum space $\widetilde{f}_A({\bf k}^2)$ is a 
smooth function of ${\bf k}^2$.
\end{thebibliography}
\end{document}